\documentclass[prb,
  reprint,
  superscriptaddress,
  amsmath,amssymb,
  floatfix,
]{revtex4-2}

\usepackage{graphicx}
\graphicspath{{figures/pcfc/}}
\usepackage{dcolumn}
\usepackage{bm}
\usepackage{xcolor}
\usepackage{booktabs}
\usepackage{multirow}
\usepackage{tikz}
\usetikzlibrary{positioning, fit, arrows.meta, shapes.geometric, shapes.misc}
\usepackage{hyperref}

\begin{document}

\title{\textit{Ab initio} High-Pressure Phase Diagrams of Al--Mg Alloys
  in the Low Solute Concentration Limit}

\author{Shambhu Bhandari Sharma}
\email{shambhu.sharma.22@ucl.ac.uk}
\affiliation{Department of Earth Sciences, University College London,
  Gower Street, London WC1E 6BT, United Kingdom}

\author{Shailesh Mehta}
\affiliation{AWE Nuclear Security Technologies, Aldermaston,
  Reading, Berkshire RG7 4PR, United Kingdom}

\author{Dario Alf\`e}
\affiliation{Department of Earth Sciences, University College London,
  Gower Street, London WC1E 6BT, United Kingdom}
\affiliation{London Centre for Nanotechnology, University College London,
  London WC1E 6BT, United Kingdom}
\affiliation{Dipartimento di Fisica Ettore Pancini, Universit\`a di Napoli
  Federico II, Monte S.\ Angelo, Napoli I-80126, Italy}

\date{\today}

\begin{abstract}
Binary alloy phase diagrams at high pressure are essential for understanding 
solidification and chemical partitioning in both engineered materials and planetary 
interiors, yet their experimental determination becomes increasingly challenging 
under extreme conditions. We apply a fully \textit{ab initio} approach 
[{\href{https://pubs.aip.org/aip/jcp/article/162/18/184502/3346477/}{J. Chem. Phys. 162, 184502 (2025)}}], based on density-functional theory, to compute 
the dilute-limit phase diagram of the Al--Mg system from ambient conditions up to 
150~GPa. As a first step, we calculate the melting curves of pure fcc Al and pure 
hcp/bcc Mg, including the hcp--bcc phase boundary and triple point of Mg, all of which agree 
closely with available experimental data. The binary phase diagram is then 
constructed at both compositional extremes. On the Al-rich side, Mg consistently 
favours the liquid throughout the entire pressure range, with this preference 
strengthening monotonically under compression. On the Mg-rich side, Al partitions preferentially into the liquid at 
ambient pressure but undergoes a complete reversal above $\sim$60~GPa, becoming 
solid-favouring as the Mg host transitions from hcp to bcc. This pressure-driven 
reversal inverts the topology of the Mg-rich coexistence field from a conventional 
downward-sloping to an upward-sloping phase boundary, and has direct implications for 
models of planetary differentiation and interior chemical stratification.
\end{abstract}

\maketitle
 
\section{Introduction}

Phase diagrams map the equilibrium stability of competing phases as functions of
temperature, pressure, and composition, and are the primary thermodynamic guide for
understanding solidification, partitioning, and chemical evolution of materials.
Even the simplest isomorphous binary metallic phase diagram already comprises distinct phase fields,
including single-phase liquid and solid-solution regions and two-phase coexistence
regions bounded by the solidus and liquidus. The solidus and liquidus boundaries,
together with the solute partition coefficient, govern macrosegregation during
solidification and the precipitation and dissolution of secondary phases during
heat treatment during processing of alloys~\cite{gaskell2024introduction,okamoto2016alloy}. In planetary
science, the same thermodynamic framework, extended to high pressure, determines
how elements fractionate between solid and liquid phases during core solidification
and mantle crystallisation, controlling the long-term chemical stratification of
planetary interiors~\cite{gaskell2024introduction,okamoto2016alloy}. Phase diagrams
are therefore the fundamental thermodynamic tool across materials science,
metallurgy, and geoscience~\cite{SchonJansen2009,ISIJReview2023}.

Phase diagram determination has been pursued by a broad spectrum of approaches.
Classical experimental methods, including equilibrated-alloy analysis, diffusion
couples, and thermal analysis techniques such as differential scanning calorimetry,
are most effective at ambient or moderate pressures~\cite{campbell2012phase,
zhao2007methods}. The CALPHAD approach fits parametric Gibbs energy models to
experimental and computational data and extrapolates efficiently to multicomponent
systems~\cite{Lukas2007ComputationalThermodynamics,saunders1998calphad}, but its
databases are constructed almost exclusively from low-pressure data, making
extrapolation to extreme conditions unreliable when experimental input is absent.
Machine-learning methods, either as classifiers/regressors trained on thermodynamic databases
or as interatomic potential surrogates for density-functional theory (DFT), offer
high-throughput screening across composition space~\cite{iecr2022_ml_pd,
wang2024machine,mlip_phaseforge_2025}, though their reliability depends critically
on training data coverage. First-principles free-energy methods evaluate phase
boundaries directly from electronic-structure calculations without empirical
fitting, and are uniquely capable of accessing thermodynamic conditions beyond
experimental reach. For solid-state boundaries, cluster-expansion methods combined
with Monte Carlo sampling have been applied widely~\cite{van2002automating,
liu2021first}. For solid--liquid alloy equilibria,
the chemical potential and partitioning of a dilute solute in solid and liquid phases
can be evaluated using thermodynamic integration, even up to Earth's
core conditions~\cite{alfe2002ab,alfe2002composition}.

Many binary systems phase diagrams of physical and technological importance involve
one component present only at low concentration, defining the dilute-solute
regime~\cite{chipman1972thermodynamics,hirose2021light,alfe2002ab,zhang2022free}.
In this limit, the thermodynamic description simplifies considerably: the solute
chemical potential varies linearly with concentration, so the solidus and liquidus
are fully determined by the excess chemical potential of the solute in each phase,
a quantity that can be computed from first principles via thermodynamic integration
and is formally exact at zero concentration~\cite{alfe2002ab,alfe2002composition}.
In a recent paper~\cite{sharma2025ab}, we developed and validated a general
\textit{ab initio} framework for computing binary alloy phase diagrams in this
limit, applicable to any binary metallic system across wide pressure and
temperature ranges.

The Al--Mg system is of direct relevance to lightweight structural alloy design~\cite{AlMg-theory,liu2013structure} 
and, at high pressures, to the thermodynamics of Mg-rich planetary
mantles and deep interiors. Despite its physical importance, the high-pressure
binary phase diagram of Al--Mg has not previously been determined from first
principles across a wide pressure range; experimentally, binary phase boundaries
are well characterised only at ambient pressure~\cite{murray1982mg}. Here we apply the framework of
Ref.~\cite{sharma2025ab} to the Al--Mg system from ambient conditions up to
150~GPa, establishing the \textit{ab initio} melting curves of pure Al and Mg,
the hcp--bcc solid-state transition and triple point of Mg, and the dilute-limit
phase boundary at both compositional extremes. 
\section{Methodology}

\begin{figure}[t]
\centering
\includegraphics[width=\columnwidth]{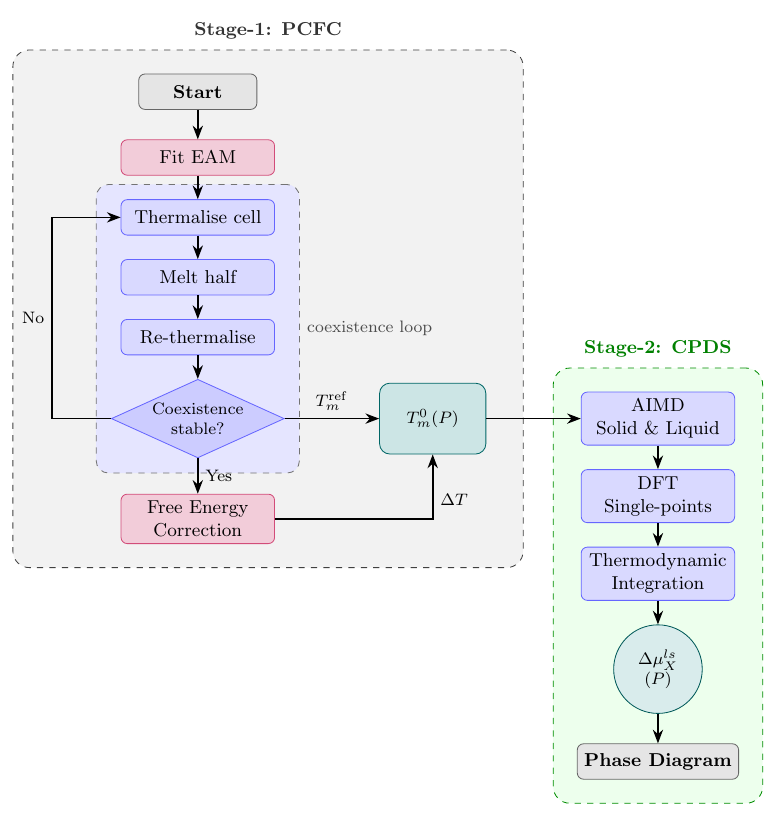}
\caption{Computational workflow for computing the \textit{ab initio} binary
alloy phase diagram in the dilute-solute limit. Stage~1 (PCFC) establishes
the pure-solvent melting curve $T_m^0(P)$ 
via large-scale coexistence simulations and free-energy correction
to DFT accuracy. Stage~2 (CPDS) computes the solute excess chemical potential
difference $\Delta\mu_X^{ls}(P)$ via \textit{ab initio} molecular dynamics and
perturbative thermodynamic integration. The two outputs are combined through
dilute-solution thermodynamic relations to construct the solidus and liquidus.}
\label{fig:workflow}
\end{figure}

The theoretical framework used in this work is described in full detail in
Ref.~\cite{sharma2025ab}; here we summarise the computational approach and the
key thermodynamic relations for computing binary alloy phase diagrams in the
dilute-solute limit. The framework proceeds in two stages. In the first, the
melting curve of the pure solvent is established via the phase coexistence with
free-energy correction (PCFC) approach. In the second, the solute excess chemical
potential difference between the solid and liquid phases is computed via the
chemical potential difference of solute (CPDS) approach. These quantities are
then combined through dilute-solution phase-equilibrium relations to construct
the solidus and liquidus of the binary alloy. The full workflow is illustrated
in Fig.~\ref{fig:workflow}.

\subsection{Pure-Solvent Melting Curve: PCFC}

The melting curve of each pure component is established using the PCFC approach.
The reference potential based on the embedded-atom method (EAM)~\cite{daw1983semiempirical} is parameterised to DFT
energies and pressures from AIMD trajectories of the solid and liquid phases.
Since any EAM potential trained at a given pressure degrades in transferability
when applied far from its training conditions, a multi-potential strategy
is adopted: two independent EAM potentials are trained per crystal structure 
at distinct pressures and applied within their respective regimes.

The
reference potential drives large-scale classical molecular dynamics (CMD)
coexistence simulations, in which a simulation cell is prepared with solid and liquid regions in contact, and evolved in the
microcanonical (NVE) ensemble until a stationary solid--liquid interface is
established. The time-averaged temperature and
pressure of this stationary state define the reference melting point $(P,
T_m^{\mathrm{ref}})$, with statistical uncertainties estimated from a reblocking
analysis~\cite{AllenTildesley2017}.

The reference melting temperature is then corrected to DFT accuracy via a
perturbative expansion of the Gibbs free energy difference between the DFT and
EAM Hamiltonians. To first order, the temperature shift $\Delta T =
T_m^{\mathrm{DFT}} - T_m^{\mathrm{ref}}$,
\begin{equation}
    \Delta T \approx \frac{\Delta G^{ls}(T_m^{\mathrm{ref}})}{S_{\mathrm{ref}}^{ls}},
\end{equation}
where $S_{\mathrm{ref}}^{ls}$ is the entropy of fusion of the reference system, and $\Delta G^{ls} = \Delta G^l - \Delta G^s$ is the liquid--solid difference in
the Gibbs free energy correction. The Helmholtz free energy difference between the
two Hamiltonians is first evaluated perturbatively from DFT single-point energies on
configurations sampled from the EAM ensemble,
\begin{equation}
    \Delta F \approx \langle \Delta U \rangle_{\mathrm{ref}}
    - \frac{\langle \delta\Delta U^2 \rangle_{\mathrm{ref}}}{2k_{\mathrm{B}}T},
\end{equation}
where $\Delta U = U_{\mathrm{DFT}} - U_{\mathrm{ref}}$ and $\delta\Delta U =
\Delta U - \langle\Delta U\rangle_{\mathrm{ref}}$. The Gibbs free energy
correction is then obtained by incorporating the induced pressure change
$\Delta p = p_{\mathrm{DFT}} - p_{\mathrm{ref}}$,
\begin{equation}
    \Delta G \approx \Delta F - \frac{V \Delta p^2}{2K_T},
\end{equation}
where $K_T$ is the isothermal bulk modulus~\cite{alfe2002complementary}.

\paragraph{Solid-State Phase Transition}

When a pure component undergoes a pressure-induced structural transition, the
stable solid phase at each pressure must be identified before the melting curve
can be constructed, since it determines which crystal structure serves as the host
lattice in the subsequent CPDS calculations. The stable phase at any $(P, T)$ is
the one with the lowest Gibbs free energy, and the phase boundary is located by
the condition $G_\alpha(P,T) = G_\beta(P,T)$. Following Mehta \textit{et
al.}~\cite{mehta2006ab}, the Gibbs free energy of each competing structure is
computed from DFT within the quasiharmonic approximation~\cite{alfe2009phon},
with the electronic contribution parametrised by a Birch--Murnaghan equation of
state and the ionic contribution by an effective Einstein model. Solving the
equal-free-energy condition numerically yields the transition
pressure and traces the full phase boundary.

\subsection{Solute Chemical Potential and Phase Diagram: CPDS}

The central quantity for constructing the binary phase diagram in the dilute-solute
limit is the excess chemical potential difference $\Delta\mu_X^{ls} =
\mu_X^{\dagger l} - \mu_X^{\dagger s}$, which fully determines both the solute
partition coefficient and the solidus and liquidus boundaries. In the limit of
vanishing solute concentration $c_X \to 0$, the solute chemical potential takes
the form~\cite{alfe2002ab,alfe2002composition}
\begin{equation}
    \mu_X = k_{\mathrm{B}}T \ln c_X + \mu_X^\dagger + \mathcal{O}(c_X),
\end{equation}
where $\mu_X^\dagger$ is the excess chemical potential in the limit of zero concentration.
 Imposing solid--liquid coexistence,
$\mu_X^s = \mu_X^l$, yields to leading order the partition coefficient
\begin{equation}
    k = \frac{c_X^s}{c_X^l} =
    \exp\!\left[\frac{\Delta\mu_X^{ls}}{k_{\mathrm{B}}T_m}\right].
    \label{eq:k}
\end{equation}
The shift in melting temperature with solute concentration, relative to the pure-solvent value $T_m^0 \equiv T_m^{\mathrm{DFT}}$, defines the liquidus and solidus curves.
\begin{align}
    T_m^l &= T_m^0 + \frac{k_{\mathrm{B}}T_m^l}{\Delta s_A^0}
    \left(e^{\Delta\mu_X^{ls}/k_{\mathrm{B}}T_m^l} - 1\right) c_X^l,
    \label{eq:liquidus}\\
    T_m^s &= T_m^0 + \frac{k_{\mathrm{B}}T_m^s}{\Delta s_A^0}
    \left(1 - e^{-\Delta\mu_X^{ls}/k_{\mathrm{B}}T_m^s}\right) c_X^s,
    \label{eq:solidus}
\end{align}
where $T_m^0(P)$ and $\Delta s_A^0(P)$ are the pure-solvent melting temperature
and entropy of fusion. Equations~\eqref{eq:liquidus}
and~\eqref{eq:solidus} are solved self-consistently at each pressure to yield the
phase diagram over the composition range in which the dilute approximation remains
valid.

\paragraph{Computing $\Delta\mu_X^{ls}$ via thermodynamic integration}
The excess chemical potential difference is computed by coupling the pure-solvent
system ($\lambda=0$) to the single-solute-substituted system ($\lambda=1$) through
the hybrid Hamiltonian $U_\lambda = (1-\lambda)U_0 + \lambda U_1$. The exact
Helmholtz free energy difference between the two endpoints is
\begin{equation}
    \Delta F = F_1 - F_0 = \int_0^1 \langle U_1 - U_0 \rangle_\lambda \,
    \mathrm{d}\lambda,
\end{equation}
where $\langle\cdots\rangle_\lambda$ denotes the thermal average under
$U_\lambda$~\cite{alfe2002ab}. When $U_1 - U_0$ is a weak perturbation, satisfied
here since a single solute substitution induces only a localised perturbation of
the host, the integrand $\langle\Delta U\rangle_\lambda$ is well approximated as
linear in $\lambda$, and the integral reduces to a perturbative two-endpoint
scheme. Applying second-order perturbation theory at each endpoint,
\begin{align}
    \Delta F &\approx \langle\Delta U\rangle_{\lambda=0}
    - \frac{\langle\delta\Delta U^2\rangle_{\lambda=0}}{2k_{\mathrm{B}}T},
    \label{eq:pert0}\\
    \Delta F &\approx \langle\Delta U\rangle_{\lambda=1}
    + \frac{\langle\delta\Delta U^2\rangle_{\lambda=1}}{2k_{\mathrm{B}}T},
    \label{eq:pert1}
\end{align}
where $\delta\Delta U = \Delta U - \langle\Delta U\rangle_\lambda$ is the
fluctuation of the energy difference about its mean. The two estimates in
Eqs.~\eqref{eq:pert0} and~\eqref{eq:pert1} are combined using a piecewise-linear
estimator: when the linear approximations anchored at $\lambda=0$ and $\lambda=1$
do not intersect within $[0,1]$, the integrand is effectively linear and the
simple average $\bar{U} = \tfrac{1}{2}(\langle\Delta U\rangle_0 +
\langle\Delta U\rangle_1)$ is adopted; when they intersect at an interior point
$\lambda^* \in (0,1)$, indicating curvature in the integrand, the piecewise-linear
estimator is applied, integrating each linear approximation over its respective
domain $[0,\lambda^*]$ and $[\lambda^*,1]$. 

The chemical potential difference then follows directly as $\Delta\mu_X^{ls} =
\Delta F^l - \Delta F^s$, where $\Delta F^l$ and $\Delta F^s$ are computed
independently for the liquid and solid phases.
\\
\paragraph{Implementation}
AIMD simulations in the canonical (NVT) ensemble are performed independently for
both solid and liquid phases at each $(P, T_m^0)$ state point, using two distinct
systems: the pure solvent $A$ ($\lambda=0$), and the system in which one solvent
atom is replaced by the solute $X$ ($\lambda=1$). At each $\lambda$ endpoint and for each
phase, statistically independent configurations are extracted from the AIMD
trajectory at intervals exceeding the structural correlation time. For each
extracted configuration, two DFT single-point energy evaluations are performed:
for configurations sampled from the $\lambda=0$ trajectory, the energy $U_0$ is
computed for the pure-solvent configuration and then $U_1$ is computed after
substituting one solvent atom with the solute; for configurations sampled from
the $\lambda=1$ trajectory, the procedure is reversed, computing $U_1$ for the
solute-containing configuration and $U_0$ after replacing the solute with a
solvent atom. This cross-evaluation at both endpoints is essential for computing
both $\langle\Delta U\rangle_\lambda$ and $\langle\delta\Delta
U^2\rangle_\lambda$ at each endpoint independently, without requiring any
intermediate $\lambda$ simulations.

\subsection{Computational Details}

All DFT and AIMD calculations are performed using the Vienna \textit{ab initio}
Simulation Package (VASP)~\cite{kresse1996efficient,kresse1999ultrasoft} with the
projector-augmented-wave (PAW) method~\cite{blochl1994projector} and the
Perdew--Burke--Ernzerhof (PBE) generalised gradient
approximation~\cite{perdew1996generalized}. The PAW core radii for Al and Mg are
1.01~\AA\ and 1.06~\AA\ respectively, and a plane-wave cutoff energy of 312~eV
is used throughout. Finite-temperature electronic effects are treated within the
Mermin formalism~\cite{mermin1965thermal}, and constant-temperature AIMD
simulations employ a Nos\'{e} thermostat~\cite{nose1984molecular}. Brillouin-zone
sampling uses Monkhorst--Pack grids~\cite{monkhorst1976special}, with mesh
densities converged individually for each stage of the calculation. For the
DFT single-point energy evaluations in the thermodynamic integration (CPDS stage)
symmetric
$k$-point grids up to $6\times6\times6$ are used, reduced to the irreducible
Brillouin zone of the underlying perfect crystal. This reduction is justified
because both solid and liquid phases, when averaged over hundreds of
configurations, recover approximately the full cubic symmetry of the perfect
lattice, making symmetry-reduced grids significantly more efficient at achieving
convergence with respect to the number of k-points, compared with using full
non-symmetric grids. \section{Results and Discussion}

\subsection{Pure-Component Melting Curves}
The melting temperatures $T_m^0(P)$ of pure Al and pure Mg are first determined using the PCFC
method, providing the pressure-dependent anchors for the Al-rich and Mg-rich
limits of the binary  Al-Mg  alloy phase diagram respectively.

\subsubsection{Melting Curve of fcc Al}

Aluminium crystallises in the fcc structure at ambient conditions, with a melting
point of 933~K~\cite{vocadlo2002ab}. Under compression it undergoes structural
transitions to hcp at $\sim$180~GPa and subsequently to bcc at
$\sim$370~GPa~\cite{polsin2018x,akahama2006evidence,kudasov2013lattice}, both well
beyond the pressure range of interest. The fcc phase therefore governs the melting
behaviour of Al throughout 0--150~GPa and is the only structure considered.

Two EAM reference potentials, $\mathrm{EAM}^{\mathrm{fcc}}_{\mathrm{low}}$ and
$\mathrm{EAM}^{\mathrm{fcc}}_{\mathrm{high}}$, are trained at low ($\sim$5~GPa)
and high ($\sim$100~GPa) pressure respectively; their optimised parameters are
listed in Table~\ref{tab:eam_params_all}. Using these potentials, large-scale
coexistence simulations are performed on a $10\times10\times20$ supercell of 8000
fcc Al atoms, evolved in the NVE ensemble for 300~ps with a 1~fs timestep,
confirmed by an energy drift test to be below 0.04~K/ps. Free-energy corrections are then computed from canonical-ensemble MD
simulations on 500-atom fcc supercells for both solid and liquid phases at each
reference melting point, run for 300~ps with a 1~fs timestep. At least 50
statistically independent configurations are extracted by sampling every 1000~fs
and subjected to DFT single-point calculations on a $2\times2\times2$ $k$-point
grid.

\begin{table}[t]
  \centering
  \caption{Optimised EAM parameters for all reference potentials
  used in this work. For each element and phase, low- and high-pressure potentials
  are listed. $n$ and $m$: repulsive and attractive exponents; $\epsilon$: energy
  scale (eV); $a$: length scale (\AA); $C$: dimensionless scaling factor.}
  \label{tab:eam_params_all}
  \small
  \setlength{\tabcolsep}{3pt}
  \begin{tabular*}{\columnwidth}{@{\extracolsep{\fill}}llccccc}
    \hline\hline
    System & Potential & $n$ & $m$ & $\epsilon$ & $a$ & $C$ \\
    \hline
    Al (fcc) & $\mathrm{EAM}^{\mathrm{fcc}}_{\mathrm{low}}$
              & 6.9025 & 3.6265 & 0.1558 & 3.2523 & 9.9562  \\
             & $\mathrm{EAM}^{\mathrm{fcc}}_{\mathrm{high}}$
              & 6.4638 & 2.7634 & 0.2663 & 3.0441 & 10.2388 \\
    \hline
    Mg (hcp) & $\mathrm{EAM}^{\mathrm{hcp}}_{\mathrm{low}}$
              & 7.680  & 4.105  & 0.0950 & 3.604  & 11.918  \\
             & $\mathrm{EAM}^{\mathrm{hcp}}_{\mathrm{high}}$
              & 7.965  & 5.892  & 0.0150 & 3.919  & 11.449  \\
    \hline
    Mg (bcc) & $\mathrm{EAM}^{\mathrm{bcc}}_{\mathrm{low}}$
              & 8.475  & 5.734  & 0.0666 & 3.489  & 11.906  \\
             & $\mathrm{EAM}^{\mathrm{bcc}}_{\mathrm{high}}$
              & 8.529  & 5.909  & 0.0111 & 3.875  & 11.497  \\
    \hline\hline
  \end{tabular*}
\end{table}

\begin{figure*}[ht]
  \centering
  \includegraphics[width=\linewidth]{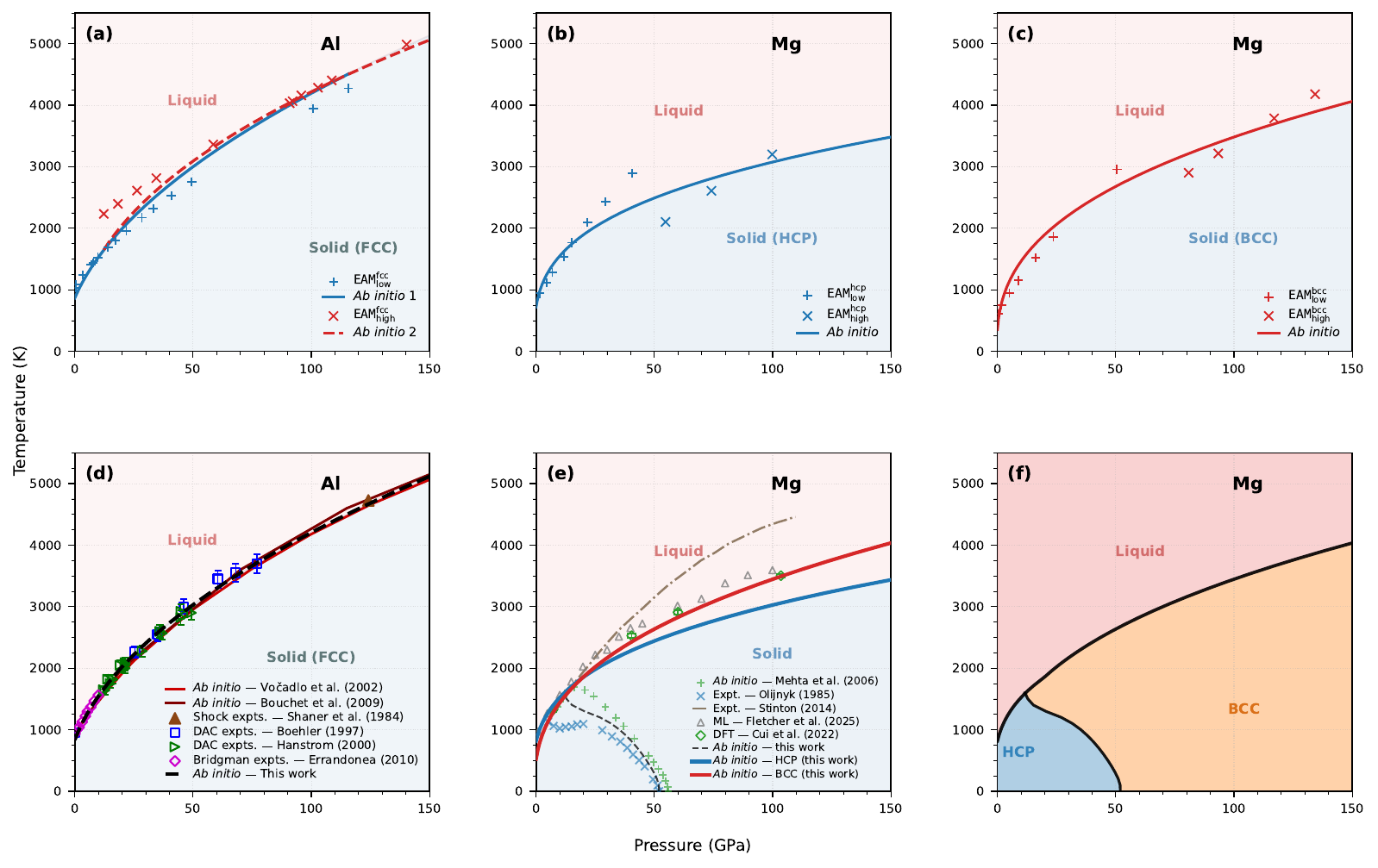}
  \caption{\textit{Ab initio} melting curves of fcc Al and Mg across
  0--150~GPa. (a)--(c) EAM reference melting temperatures (symbols) and
  \textit{ab initio} results (lines) for fcc Al, hcp Mg, and
  bcc Mg respectively, using low- and high-pressure reference potentials.
  (d) Composite Al melting curve compared with prior \textit{ab initio}
  calculations and experimental results. (e) Composite Mg melting curves for the hcp and bcc
  phases compared with experimental and theoretical results. (f) Computed
  Mg phase diagram showing the stability fields of hcp, bcc, and liquid
  phases and the hcp--bcc--liquid triple point.}
  \label{fig:melting_al_mg}
\end{figure*}

The EAM reference
melting temperatures and perturbatively corrected \textit{ab initio} results are
shown in Fig.~\ref{fig:melting_al_mg}(a). For
$\mathrm{EAM}^{\mathrm{fcc}}_{\mathrm{low}}$, corrections $\Delta T$ are small
at low pressure but grow increasingly as the potential moves outside its training
regime at high pressure. $\mathrm{EAM}^{\mathrm{fcc}}_{\mathrm{high}}$ shows the
complementary behaviour: corrections are largest at low pressure and decrease
steadily, becoming negligibly small at high pressure where the potential was
parametrised.  Despite starting from very different reference potentials, the two
independently corrected \textit{ab initio} melting curves converge to a mutually
consistent result over their common pressure interval ($\sim$26--115~GPa),
providing a stringent internal validation of the perturbative framework. The computed melting curve is in excellent agreement with experimental and
theoretical results  across the full pressure range~\cite{vocadlo2002ab,
bouchet2009melting,boehler1997melting,hanstrom2000high,errandonea2010melting,
homan1984high}, as shown in Fig.~\ref{fig:melting_al_mg}(d).

\subsubsection{Melting Curves of hcp and bcc Mg}

Magnesium crystallises in the hcp structure at ambient conditions and undergoes a
pressure-induced transition to bcc at $\sim$50~GPa at room
temperature~\cite{olijnyk1985high,stinton2014equation}. Since both solid phases
are present within the pressure range of interest, their melting curves and the
hcp--bcc phase boundary are established independently.

\paragraph{hcp--bcc phase boundary}
The Gibbs free energy of each phase is computed within the quasiharmonic
approximation using the finite-displacement method over a volume grid of
10.0--26.0~\AA$^3$/atom (step 0.5~\AA$^3$/atom) and temperatures of 1--6000~K
(step 100~K). Electronic energies use $32\times32\times20$ and $28\times28\times28$
$k$-point meshes for hcp and bcc respectively. Phonon calculations employ
$4\times4\times2$ (hcp) and $4\times4\times4$ (bcc) supercells, sampled with
$7\times7\times4$ and $6\times6\times6$ $k$-point meshes respectively, with
atomic displacements of 0.023~\AA\ and a $12\times12\times12$ $q$-point grid
for both phases. The phase boundary is located where the two Gibbs free energies
are equal, as shown in Fig.~\ref{fig:melting_al_mg}(e--f). The computed
transition pressure at 300~K is 49.84~GPa, in excellent agreement with the
experimental value of $50 \pm 6$~GPa~\cite{olijnyk1985high} and the
first-principles prediction of Mehta \textit{et al.}~\cite{mehta2006ab}. The
boundary carries a negative Clapeyron slope: the transition pressure decreases
with increasing temperature, reflecting the higher vibrational entropy of the
more open bcc structure, which becomes progressively stabilised relative to hcp
under heating.

\paragraph{hcp Mg melting curve}
For the hcp phase, $\mathrm{EAM}^{\mathrm{hcp}}_{\mathrm{low}}$ is trained at
low pressure ($\sim$5~GPa) and $\mathrm{EAM}^{\mathrm{hcp}}_{\mathrm{high}}$ at
high pressure ($\sim$92~GPa), providing complementary coverage of the hcp
stability field; their parameters are listed in Table~\ref{tab:eam_params_all}.
Coexistence simulations are performed on a $13\times13\times26$ supercell of 8788
Mg atoms, evolved in the NVE ensemble for 300~ps with a 1~fs timestep. Free-energy
corrections are obtained from canonical-ensemble MD simulations on 432-atom
supercells, with at least 50 statistically independent configurations extracted
by sampling every 1000~fs and evaluated with a $2\times2\times2$ $k$-point grid.
The EAM reference melting temperatures and corrected \textit{ab initio} hcp
melting curve are shown in Fig.~\ref{fig:melting_al_mg}(b). Within each
potential's fitted range, corrections $\Delta T$ are small; outside it, they grow systematically.
 The calculated hcp melting temperature at near-ambient pressure is $909
\pm 13$~K, in close agreement with the experimental value of 923~K.

\paragraph{bcc Mg melting curve}
For the bcc phase, $\mathrm{EAM}^{\mathrm{bcc}}_{\mathrm{low}}$ is trained at
intermediate pressure ($\sim$31~GPa) and
$\mathrm{EAM}^{\mathrm{bcc}}_{\mathrm{high}}$ at high pressure ($\sim$88~GPa),
consistent with the bcc stability field; their parameters are listed in
Table~\ref{tab:eam_params_all}. The coexistence simulations and free-energy
corrections follow the same protocol as for the hcp phase: a $13\times13\times26$
supercell of 8788 Mg atoms in the NVE ensemble, with 432-atom supercells for the
canonical MD corrections, $\geq$50 configurations sampled every 1000~fs, and a
$2\times2\times2$ $k$-point grid for the DFT single-point evaluations. The EAM
reference melting temperatures and corrected \textit{ab initio} bcc melting curve
are shown in Fig.~\ref{fig:melting_al_mg}(c). The bcc phase melts at only $605
\pm 64$~K near ambient pressure, well below the hcp value, reflecting the
mechanical instability of the open bcc lattice at low compression. With increasing
pressure, the bcc melting curve rises more steeply than the hcp branch and
eventually surpasses it, consistent with the progressive entropic stabilisation of
bcc under compression.

\paragraph{Complete Mg phase diagram}
The full Mg phase diagram with distinct phase fields, assembling the hcp and bcc melting curves with the
hcp--bcc phase boundary, is presented in Figs.~\ref{fig:melting_al_mg}(e)
and~\ref{fig:melting_al_mg}(f). The hcp--bcc--liquid triple point is located at
approximately $P^{\mathrm{TP}} = 11.7 $~GPa and $T^{\mathrm{TP}} = 1593$ K. As shown in
Fig.~\ref{fig:melting_al_mg}(f), below the triple point hcp Mg occupies a
compact stability field bounded above by its melting curve and to the right by the
negatively sloped hcp--bcc boundary; above it, bcc dominates the entire
high-pressure solid region up to 150~GPa. The liquid field covers the
high-temperature domain above both melting curves, with the solid-to-liquid
transition occurring via hcp below the triple point and via bcc above it.
The computed phase boundaries agree well with available experimental and
theoretical data~\cite{olijnyk1985high,mehta2006ab,cui2022melting,
fletcher2025autonomous,stinton2014equation}.

\subsection{Al--Mg Phase Diagram in the Dilute-Solute Limit}
\label{subsec:almg}
With the \textit{ab initio} melting curves of pure Al and Mg and the Mg
hcp--bcc phase boundary established, the CPDS method is applied to compute
the solidus and liquidus in the dilute-solute limit across 0--150~GPa,
covering up to 6~at.\% Mg on the Al-rich side and up to 6~at.\% Al on the
Mg-rich side.

\subsubsection{Configuration Sampling}

On the Al-rich side, AIMD simulations are performed on 108-atom
($3\times3\times3$ fcc) supercells in the NVT ensemble at four pressures along
the \textit{ab initio} Al melting curve: 0, 54, 100, and 149~GPa, at the
corresponding melting temperatures of 930, 3204, 4277, and 5056~K. On the
Mg-rich side, hcp supercells ($5\times5\times5$, 250 atoms) are used at 0~GPa
and bcc supercells ($6\times6\times6$, 216 atoms) at 60, 90, and 150~GPa,
consistent with the calculated hcp--bcc phase boundary, at the corresponding
\textit{ab initio} Mg melting temperatures of 920, 2823, 3306, and 4037~K. In
both cases, independent trajectories are generated for the pure solvent
($\lambda=0$) and the single-solute-substituted system ($\lambda=1$) in both
solid and liquid phases. The thermodynamic inputs extracted from the AIMD trajectories are summarised in
Table~\ref{tab:thermo_inputs}. For Al, $\Delta V = V_l - V_s$ decreases from
1.29~\AA$^3$/atom at 0~GPa to 0.19~\AA$^3$/atom at 149~GPa (a factor of
$\sim$7), and $\Delta s_{\mathrm{Al}}^0$ decreases smoothly from 1.30 to
0.88~$k_{\mathrm{B}}$/atom, reflecting the progressive structural convergence of
solid and liquid under compression. For Mg, the compression of $\Delta V$ is far
more pronounced, from 1.15 to 0.06~\AA$^3$/atom (a factor of $\sim$19),
consistent with the considerably higher compressibility of Mg relative to Al.
Notably, $\Delta s_{\mathrm{Mg}}^0$ exhibits a near-plateau between 60 and
90~GPa (0.82 to 0.81~$k_{\mathrm{B}}$/atom), departing from the otherwise
smooth monotonic decrease. This coincides with the bcc stability field: the more
open and less ordered bcc lattice is structurally closer to the liquid than hcp,
partially offsetting the compression-driven convergence of the two phases.

\begin{table}[t]
  \centering
  \caption{Thermodynamic properties of the equilibrated solid and liquid phases
  along the \textit{ab initio} melting curve, obtained from the AIMD trajectories
  used in the thermodynamic integration. For Mg, the solid phase is hcp at 0~GPa
  and bcc at 60--150~GPa. All volumes in \AA$^3$/atom; $\Delta V = V_l - V_s$;
  $\Delta s_A^0$ in $k_{\mathrm{B}}$/atom.
  $^a$DFT (GGA-PBE)~\cite{vocadlo2002ab,hong2019reentrant};
  $^b$experiment~\cite{sansonetti2005handbook,cannon1974behavior,chase1985janaf,
  courac2020thermoelastic,chase1998nist}.}
  \label{tab:thermo_inputs}
  \small
  \begin{tabular*}{\columnwidth}{@{\extracolsep{\fill}}llcrrrr}
    \hline\hline
    System & $P$ (GPa) & $T$ (K) & $V_s$ & $V_l$ & $\Delta V$ & $\Delta s_A^0$ \\
    \hline
    Al          &   0 &  930 & 17.78 & 19.07 & 1.29 & 1.30 \\
    Ref.$^a$    &   0 &  912 & 17.70 & --    & 1.51 & 1.35 \\
    Ref.$^b$    &   0 &  933 & --    & --    & 1.24 & 1.38 \\
                &  54 & 3204 & 12.63 & 13.03 & 0.40 & 0.97 \\
                & 100 & 4277 & 10.94 & 11.20 & 0.26 & 0.92 \\
                & 149 & 5056 &  9.82 & 10.02 & 0.19 & 0.88 \\
    \hline
    Mg          &   0 &  920 & 24.76 & 25.91 & 1.15 & 0.99 \\
    Ref.$^a$    &   0 &  890 & 24.30 & 25.40 & 1.10 & --   \\
    Ref.$^b$    &   0 &  923 & 24.65 & 25.38 & 0.76 & 1.11 \\
                &  60 & 2823 & 13.52 & 13.71 & 0.20 & 0.82 \\
                &  90 & 3306 & 11.99 & 12.13 & 0.13 & 0.81 \\
                & 150 & 4037 & 10.18 & 10.24 & 0.06 & 0.77 \\
    \hline\hline
  \end{tabular*}
\end{table}

\begin{figure*}[tbp]
  \centering
  \includegraphics[width=\linewidth]{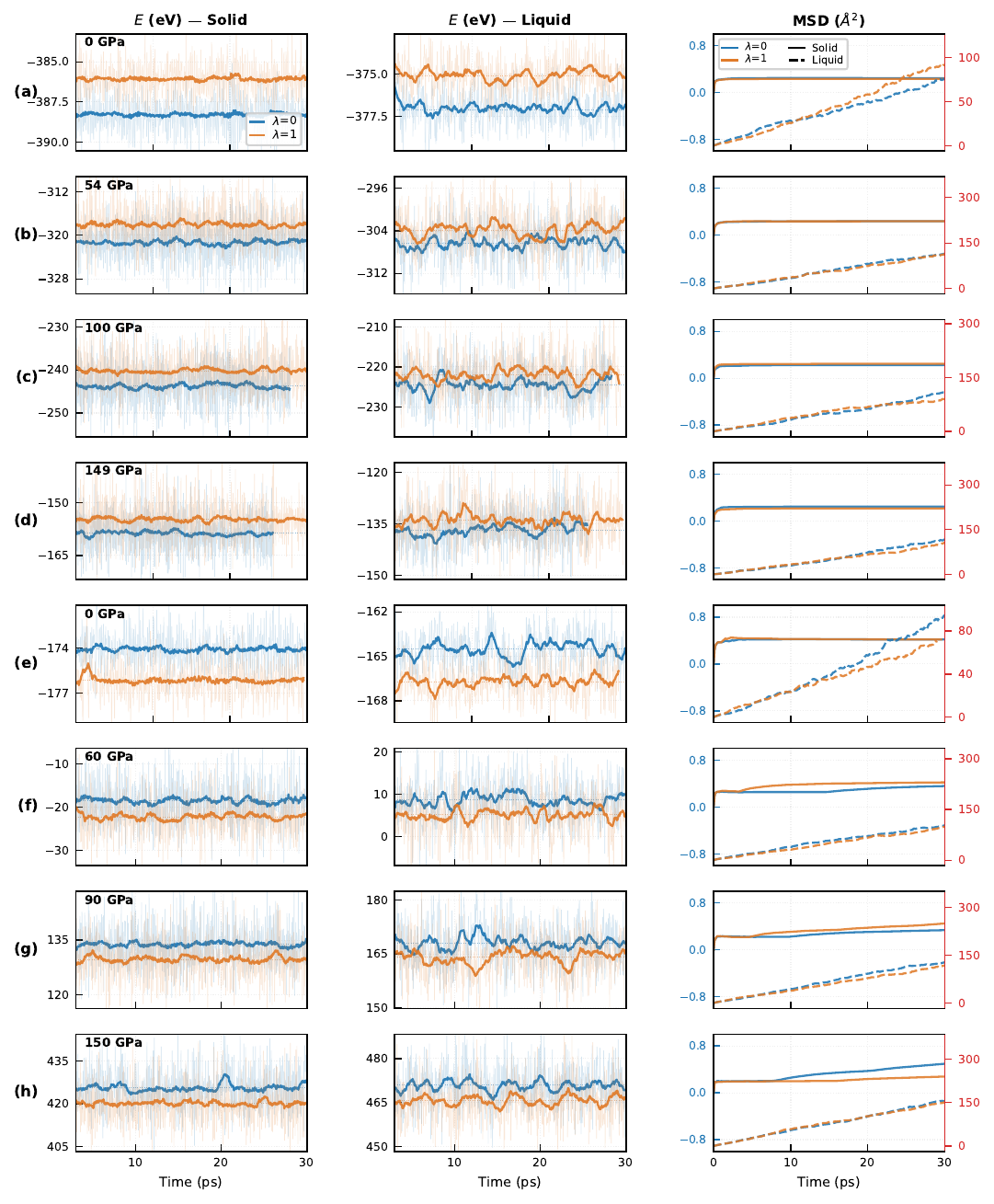}
  \caption{AIMD trajectory diagnostics at each pressure. Rows (a)--(d): Al-rich
  limit (Mg solute in Al) at 0, 54, 100, and 149~GPa. Rows (e)--(h): Mg-rich
  limit (Al solute in Mg) at 0, 60, 90, and 150~GPa. Each row shows the
  instantaneous energy for $\lambda=0$ (blue) and $\lambda=1$ (orange)
  in the solid (left) and liquid (centre) phases, with rolling averages as thick
  lines; the right panel shows the MSD for solid (blue, left axis) and liquid
  (red, right axis), with solid and dashed lines for $\lambda=0$ and $\lambda=1$
  respectively. In rows (a)--(d) the $\lambda=1$ energy lies consistently above
  $\lambda=0$; in rows (e)--(h) the ordering is reversed, reflecting the stronger
  bonding of Al in the Mg host.}
  \label{fig:ti_diagnostics}
\end{figure*}

The trajectory diagnostics for both limits are shown in
Fig.~\ref{fig:ti_diagnostics}. In both cases, the
energy traces for $\lambda=0$ and $\lambda=1$ fluctuate about
well-defined stationary means with no systematic drift, confirming stable NVT
sampling throughout. The mean-square displacement (MSD) curves confirm phase identity at all pressures:
solid trajectories plateau rapidly to $\sim$0.2~\AA$^2$, while liquid
trajectories grow linearly, confirming sustained diffusion. The MSD curves for
$\lambda=0$ and $\lambda=1$ are nearly indistinguishable in both phases at all
pressures, confirming that the single solute substitution does not structurally
or dynamically perturb the host, consistent with the weak-perturbation
requirement of the thermodynamic integration scheme.

A physically important contrast between the two limits is already visible at this
stage. On the Al-rich side (\ref{fig:ti_diagnostics}a--d), the $\lambda=1$ energy
lies consistently \emph{above} $\lambda=0$ in both phases at all pressures,
indicating that substituting Mg into Al raises the total energy. On the Mg-rich
side (\ref{fig:ti_diagnostics}e--h), the ordering is \emph{reversed}: $\lambda=1$
lies consistently \emph{below} $\lambda=0$, reflecting the stronger bonding of
Al relative to Mg in the Mg host. This reversed energy offset is the direct
microscopic precursor of the partitioning reversal on the Mg-rich side, as
quantified in the following section.

\subsubsection{Chemical Potential Estimation}

\begin{figure*}[tbp]
  \centering
  \includegraphics[width=\linewidth]{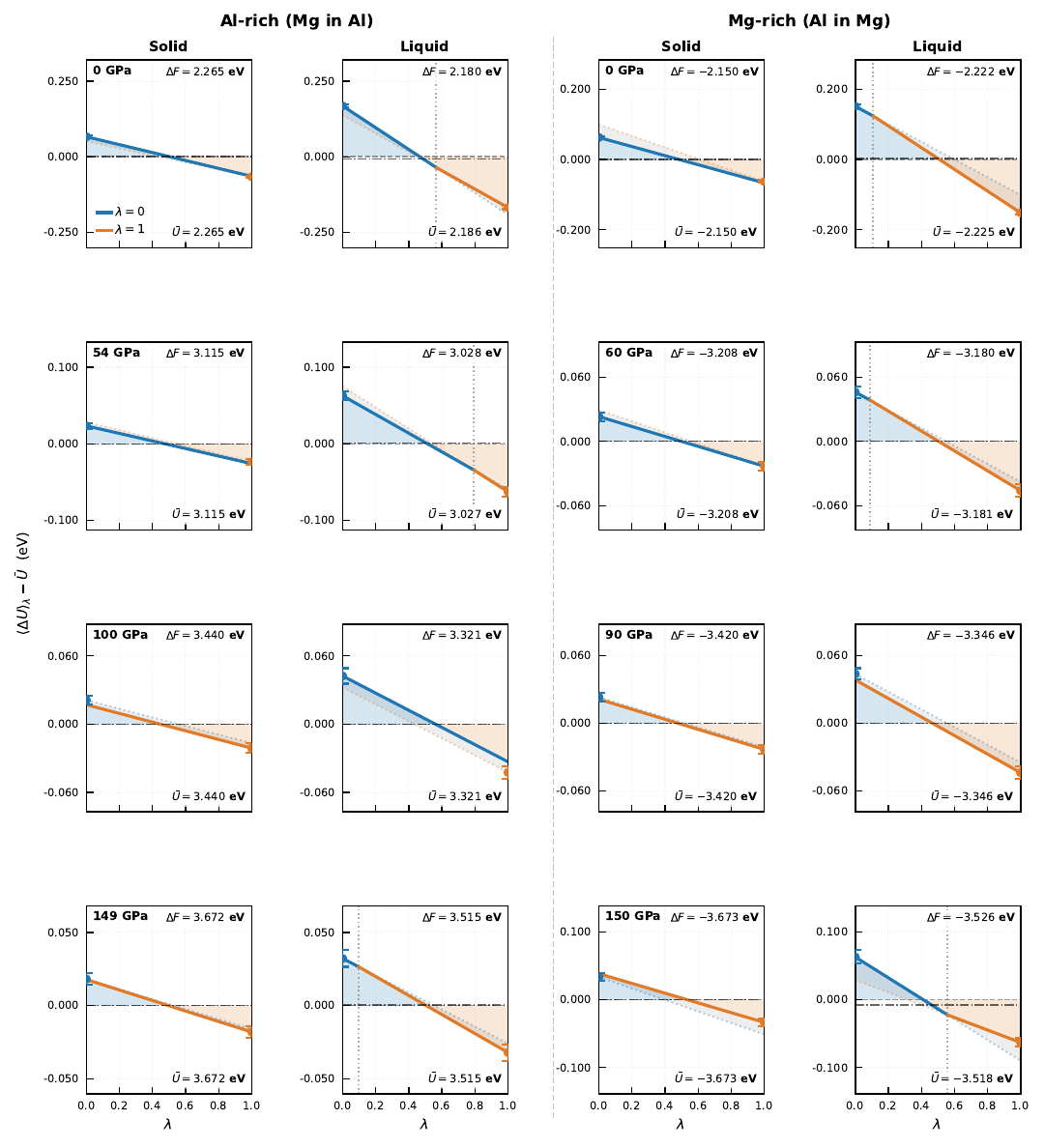}
  \caption{Perturbative TI integrand $\langle\Delta U\rangle_\lambda - \bar{U}$
  as a function of $\lambda$ for the Al-rich limit (Mg solute in Al, left two
  columns) and the Mg-rich limit (Al solute in Mg, right two columns), each
  showing solid and liquid phases at four pressures (rows). Subtracting
  $\bar{U}$ centres the ordinate so that deviations from zero
  directly quantify non-linearity. Blue and orange lines show the linear
  approximations anchored at $\lambda=0$ and $\lambda=1$ respectively; shaded
  bands give the $\pm1\sigma$ uncertainty corridor.}
  \label{fig:ti_integration}
\end{figure*}

The validity of the two-endpoint perturbative scheme is first assessed through
the TI integrand $\langle\Delta U\rangle_\lambda - \bar{U}$, shown as a function
of $\lambda$ for both compositional limits in Fig.~\ref{fig:ti_integration}.
Subtracting the simple average $\bar{U}$ centres the ordinate so that a perfectly
linear integrand appears as two coincident lines passing through zero; any
departure from zero directly measures non-linearity. In both limits, in both
phases, and at all pressures, the blue ($\lambda=0$) and orange ($\lambda=1$)
linear approximations are in close mutual agreement, the $\pm1\sigma$ uncertainty
corridor between them is extremely narrow, and the fluctuation widths
$\sqrt{\langle\delta\Delta U^2\rangle}$ are small. This confirms that both the
Al$\to$Mg and Mg$\to$Al transmutations constitute weak perturbations and that
$\langle\Delta U\rangle_\lambda$ is well approximated as linear across $[0,1]$,
validating the two-endpoint integration scheme throughout.

The coloured fills provide an additional visual indicator of which estimator is
applied. Where the two linear approximations do not intersect within $[0,1]$, the
blue and orange fills are approximately equal in area and opposite in sign,
confirming that the correction to $\bar{U}$ is negligible and the simple-average
estimator is adopted. Where the approximations cross at an
interior point $\lambda^* \in (0,1)$, the fills are unequal and encode a non-zero
signed correction, so the piecewise-linear estimator is applied instead. In all
cases the correction beyond $\bar{U}$ is small, confirming
that the perturbative assumption is well satisfied across both limits and all
pressures.

\begin{figure}[tbp]
  \centering
  \includegraphics[width=\columnwidth]{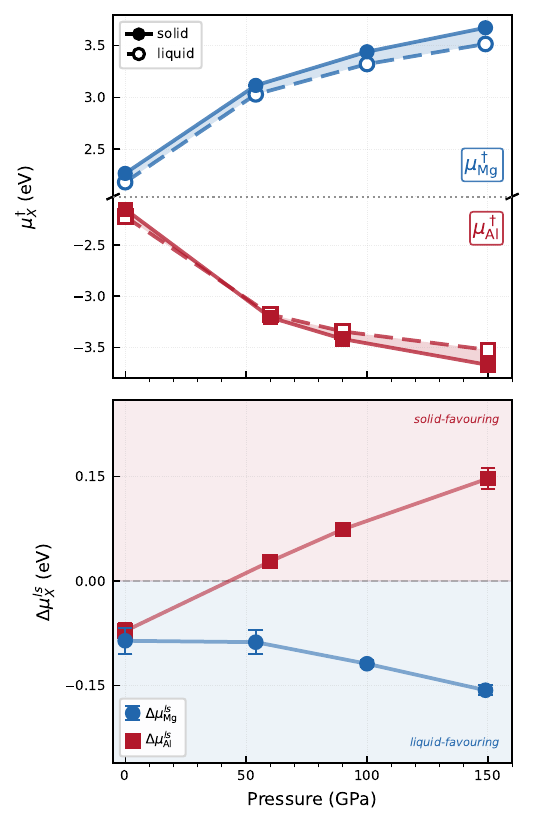}
  \caption{Pressure evolution of the excess chemical potentials and their
  difference for Mg in Al (blue) and Al in Mg (red). \emph{Upper panel}:
  individual solid ($\mu_X^{\dagger s}$, filled symbols) and liquid
  ($\mu_X^{\dagger l}$, open symbols) excess chemical potentials; the shaded
  band between each solid--liquid pair highlights the sign and magnitude of the
  partitioning tendency. }
  \label{fig:dmu_k_summary}
\end{figure}

The resulting excess chemical potentials are shown in
Fig.~\ref{fig:dmu_k_summary}. The upper panel reveals a striking mirror
symmetry between the two limits: $\mu_{\mathrm{Mg}}^{\dagger s}$ and
$\mu_{\mathrm{Mg}}^{\dagger l}$ are both positive and rise monotonically with
pressure, while $\mu_{\mathrm{Al}}^{\dagger s}$ and $\mu_{\mathrm{Al}}^{\dagger
l}$ are both negative and decrease monotonically, with comparable magnitudes
across the full pressure range. This near-antisymmetry reflects the close chemical
affinity of Al and Mg as neighbouring elements: substituting Mg into Al costs
energy, raising the excess chemical potential above zero, while substituting Al
into Mg releases energy by a similar amount, driving it below zero. The shaded
band between each solid--liquid pair encodes the partitioning tendency: a narrow
band indicates weak fractionation, while a wide band indicates strong partitioning.
On the Al-rich side, the solid value of $\mu_{\mathrm{Mg}}^\dagger$ lies
consistently above the liquid value at all pressures, and the gap between them
widens progressively with compression. On the Mg-rich side, the two values
converge with increasing pressure, cross, and reverse their ordering above
$\sim$60~GPa, so that the solid ultimately lies below the liquid.

Despite the overall mirror symmetry in magnitude, it is precisely this reversal
in the solid--liquid ordering on the Mg-rich side that breaks the symmetry
between the two limits. The lower panel of Fig.~\ref{fig:dmu_k_summary} makes
this explicit by showing $\Delta\mu_X^{ls} = \mu_X^{\dagger l} -
\mu_X^{\dagger s}$ directly, with blue shading for liquid-favouring
($\Delta\mu_X^{ls} < 0$) and red shading for solid-favouring ($\Delta\mu_X^{ls}
> 0$) regimes.

On the Al-rich side, $\Delta\mu_{\mathrm{Mg}}^{ls}$ grows
monotonically more negative with pressure, reflecting a continuously strengthening
liquid-favouring preference for Mg with no sign of reversal across the full
pressure range. On the Mg-rich side, $\Delta\mu_{\mathrm{Al}}^{ls}$ begins
negative at ambient pressure, crosses zero around 50~GPa, and grows
steadily positive thereafter without saturation. This sign reversal signals a complete change in the direction
of Al partitioning under compression, structurally gated by the hcp--bcc
transition of the Mg host, and directly produces the topological inversion of the
Mg-rich phase diagram discussed in the following section.

\subsubsection{Phase Diagram}

With $T_m^0(P)$, $\Delta\mu_X^{ls}(P)$ and $\Delta s_A^0(P)$ established, the solidus and
liquidus at both compositional extremes are constructed by solving the
dilute-solution phase-equilibrium relations (eq. \ref{eq:solidus}, and \ref{eq:liquidus}) self-consistently at each pressure.

\begin{figure*}[tbp]
  \centering
  \includegraphics[width=\linewidth]{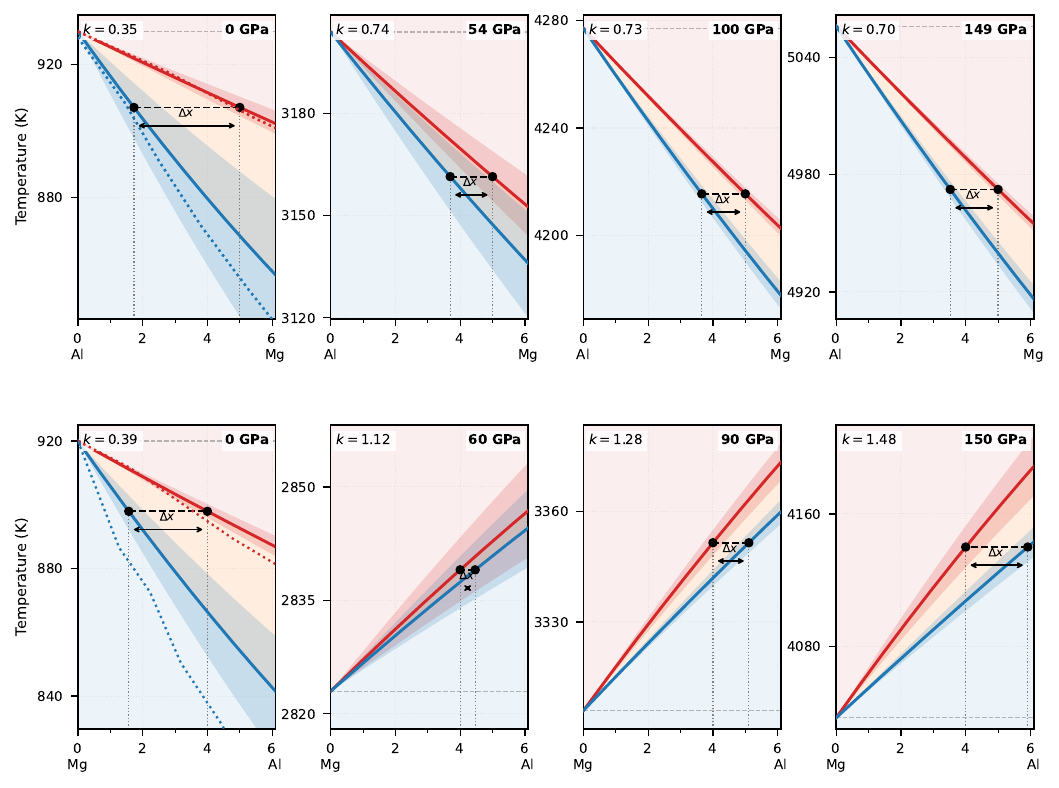}
  \caption{\textit{Ab initio} Al--Mg dilute-solute phase diagrams. Top row
  (a)--(d): Al-rich limit (Mg solute in Al) at 0, 54, 100, and 149~GPa. Bottom
  row (e)--(h): Mg-rich limit (Al solute in Mg) at 0, 60, 90, and 150~GPa.
  Liquidus (red) and solidus (blue) are shown with one-sigma uncertainty bands;
  orange shading marks the two-phase coexistence field. Representative tie lines
  at $x_{\mathrm{Mg}}=5\%$ (top) and $x_{\mathrm{Al}}=4\%$ (bottom)
  illustrate $k=c_s/c_l$ and the coexistence width $\Delta x$.}
  \label{fig:phase_diagram}
\end{figure*}

On the Al-rich side, the resulting phase diagrams are shown in the top row of
Fig.~\ref{fig:phase_diagram}. At ambient pressure, the computed solidus
and liquidus reproduce the experimental data of Murray~\cite{murray1982mg}
closely across the full 0--6~at.\% Mg range, providing direct validation of the framework. The partition
coefficient $k = 0.35$ indicates that the solid retains only about one third of
the Mg present in the coexisting liquid, producing a correspondingly wide
coexistence field. As pressure increases to 54~GPa, $k$ rises sharply to 0.74
and the coexistence width narrows substantially. This narrowing is driven
primarily by the reduction in $\Delta s_{\mathrm{Al}}^0$ from
$1.30\,k_{\mathrm{B}}$/atom at ambient to $0.97\,k_{\mathrm{B}}$/atom at
54~GPa: as the entropy of fusion decreases, the liquidus and solidus slopes
steepen and converge, compressing the two-phase field even as
$|\Delta\mu_{\mathrm{Mg}}^{ls}|$ grows. At 100 and 149~GPa, $k$ stabilises near
0.73 and 0.70 respectively, indicating that the competing effects of a growing
$|\Delta\mu_{\mathrm{Mg}}^{ls}|$ and a declining $\Delta s_{\mathrm{Al}}^0$
increasingly balance, driving $k$ toward a high-pressure asymptote. The overall
topology is preserved throughout: both boundaries slope negatively with Mg
concentration and Mg consistently favours the liquid across the full pressure
range.

On the Mg-rich side, shown in the bottom row of Fig.~\ref{fig:phase_diagram}, the pressure
evolution is qualitatively richer. At ambient pressure, the framework reproduces
the experimental data of Murray~\cite{murray1982mg} closely, with $k = 0.39$ and
the liquidus slope agreeing to within $\sim$6\%, validating the approach on the
Mg-rich side. At 60~GPa, the sign reversal in $\Delta\mu_{\mathrm{Al}}^{ls}$ has
already taken full effect: the topology of the diagram inverts, with both
boundaries now sloping positively with Al concentration and $k$ rising to 1.12.
The solid now incorporates more Al than the coexisting liquid, and the melting
temperature increases with Al addition rather than decreasing. The coexistence
field is relatively narrow at 60~GPa, reflecting the modest magnitude of
$\Delta\mu_{\mathrm{Al}}^{ls}$ just above the reversal pressure combined with the
reduced $\Delta s_{\mathrm{Mg}}^0$. At 90~GPa, solid-favouring partitioning is
more firmly established with $k = 1.28$, and the coexistence field begins to
re-broaden as the growing magnitude of $\Delta\mu_{\mathrm{Al}}^{ls}$
increasingly offsets the entropy-driven narrowing. By 150~GPa, the reversal is
fully developed: $k = 1.48$ and the coexistence field has re-broadened
substantially. Unlike the Al-rich side where $k$ saturates, $k$ on the Mg-rich
side continues to grow without bound with increasing pressure, reflecting the
continuously strengthening structural preference of Al for the bcc solid.

The two limits together reveal a fundamental asymmetry in how solute
partitioning evolves under extreme compression: on the Al-rich side, the
liquid-favouring preference of Mg strengthens monotonically and saturates at high
pressure, driven by the growing misfit penalty of the oversized Mg atom in the
increasingly dense melt; on the Mg-rich side, the partitioning of Al undergoes a
complete reversal, driven not by compression alone but by the structural
reorganisation of the host from hcp to bcc, which opens larger and more
geometrically tolerant substitution sites that progressively favour the solid over
the liquid. This structural gating of the partitioning reversal, absent on the
Al-rich side where no such transition occurs, is what distinguishes the two limits
fundamentally and makes the Mg-rich phase diagram qualitatively richer under
compression.

 \section{Conclusion}
\label{sec:conclusion}

We have applied a fully \textit{ab initio} framework to compute the dilute-limit
phase diagram of the Al--Mg binary system from ambient conditions up to 150~GPa,
establishing melting curves for pure fcc Al and hcp/bcc Mg, the Mg hcp--bcc phase
boundary and triple point, and the solidus, liquidus, and solute partition
coefficient at both compositional extremes.

The \textit{ab initio} melting curves of both metals agree closely with available
experimental and theoretical data across the full pressure range, validating the
PCFC approach in two structurally distinct systems. For Mg, the hcp--bcc
transition requires independent treatment of the two structural branches; the
composite melting curve is thermodynamically continuous across the transition and
the computed triple point is in reasonable agreement with experimental estimates.

On the Al-rich side, Mg consistently favours the liquid throughout 0--150~GPa,
with the liquid-favouring preference strengthening monotonically under compression
and the partition coefficient approaching a high-pressure asymptote. The computed
phase boundaries reproduce the experimental data of Murray~\cite{murray1982mg}
closely at ambient pressure, validating the framework on this side.

On the Mg-rich side, Al partitions into the liquid at ambient pressure but
undergoes a complete pressure-driven reversal above $\sim$60~GPa, becoming
solid-favouring as the Mg host transitions from hcp to bcc. The partition
coefficient crosses unity in this pressure interval and continues to rise without
saturation. This reversal inverts the topology of the Mg-rich coexistence field
from a conventional downward-sloping to an upward-sloping phase boundary, a
feature with no counterpart on the Al-rich side. Together, the two limits reveal
a fundamental thermodynamic asymmetry under compression that is structurally gated
by the hcp--bcc transition of the Mg host.

The present results have direct implications for models of planetary
differentiation and interior chemical stratification. The pressure-driven reversal
of Al partitioning above $\sim$60~GPa implies that in Mg-rich planetary interiors,
Al would be preferentially incorporated into the growing solid during progressive
crystallisation rather than expelled into the residual melt, a fractionation
scenario qualitatively different from what ambient-pressure data would predict. 
\begin{acknowledgments}
We acknowledge the support of AWE Nuclear Security Technologies and the HPC facility at UCL\@.
This work also used the ARCHER2 UK National Supercomputing Service
(\url{https://www.archer2.ac.uk}), under the Mineral Consortium allocation.
\end{acknowledgments}

\section*{Author Declarations}

\subsection*{Conflict of Interest}
The authors have no conflicts to disclose.

\subsection*{Author Contributions}
\textbf{Shambhu Bhandari Sharma:} Investigation (equal); Writing -- original draft (equal).
\textbf{Shailesh Mehta:} Funding acquisition (supporting).
\textbf{Dario Alf\`{e}:} Supervision (equal).

\section*{Data Availability}
The data that support the findings of this study are available from the corresponding author
upon reasonable request.
 
\bibliography{bibliography/Introduction,bibliography/Results,bibliography/Methodology,bibliography/Conclusions}

\begin{thebibliography}{53}%
\makeatletter
\providecommand \@ifxundefined [1]{%
 \@ifx{#1\undefined}
}%
\providecommand \@ifnum [1]{%
 \ifnum #1\expandafter \@firstoftwo
 \else \expandafter \@secondoftwo
 \fi
}%
\providecommand \@ifx [1]{%
 \ifx #1\expandafter \@firstoftwo
 \else \expandafter \@secondoftwo
 \fi
}%
\providecommand \natexlab [1]{#1}%
\providecommand \enquote  [1]{``#1''}%
\providecommand \bibnamefont  [1]{#1}%
\providecommand \bibfnamefont [1]{#1}%
\providecommand \citenamefont [1]{#1}%
\providecommand \href@noop [0]{\@secondoftwo}%
\providecommand \href [0]{\begingroup \@sanitize@url \@href}%
\providecommand \@href[1]{\@@startlink{#1}\@@href}%
\providecommand \@@href[1]{\endgroup#1\@@endlink}%
\providecommand \@sanitize@url [0]{\catcode `\\12\catcode `\$12\catcode
  `\&12\catcode `\#12\catcode `\^12\catcode `\_12\catcode `\%12\relax}%
\providecommand \@@startlink[1]{}%
\providecommand \@@endlink[0]{}%
\providecommand \url  [0]{\begingroup\@sanitize@url \@url }%
\providecommand \@url [1]{\endgroup\@href {#1}{\urlprefix }}%
\providecommand \urlprefix  [0]{URL }%
\providecommand \Eprint [0]{\href }%
\providecommand \doibase [0]{https://doi.org/}%
\providecommand \selectlanguage [0]{\@gobble}%
\providecommand \bibinfo  [0]{\@secondoftwo}%
\providecommand \bibfield  [0]{\@secondoftwo}%
\providecommand \translation [1]{[#1]}%
\providecommand \BibitemOpen [0]{}%
\providecommand \bibitemStop [0]{}%
\providecommand \bibitemNoStop [0]{.\EOS\space}%
\providecommand \EOS [0]{\spacefactor3000\relax}%
\providecommand \BibitemShut  [1]{\csname bibitem#1\endcsname}%
\let\auto@bib@innerbib\@empty
\bibitem [{\citenamefont {Gaskell}\ and\ \citenamefont
  {Laughlin}(2024)}]{gaskell2024introduction}%
  \BibitemOpen
  \bibfield  {author} {\bibinfo {author} {\bibfnamefont {D.~R.}\ \bibnamefont
  {Gaskell}}\ and\ \bibinfo {author} {\bibfnamefont {D.~E.}\ \bibnamefont
  {Laughlin}},\ }\href@noop {} {\emph {\bibinfo {title} {Introduction to the
  Thermodynamics of Materials}}}\ (\bibinfo  {publisher} {CRC press},\ \bibinfo
  {year} {2024})\BibitemShut {NoStop}%
\bibitem [{\citenamefont {Okamoto}\ \emph {et~al.}(2016)\citenamefont
  {Okamoto}, \citenamefont {Schlesinger},\ and\ \citenamefont
  {Mueller}}]{okamoto2016alloy}%
  \BibitemOpen
  \bibfield  {author} {\bibinfo {author} {\bibfnamefont {H.}~\bibnamefont
  {Okamoto}}, \bibinfo {author} {\bibfnamefont {M.~E.}\ \bibnamefont
  {Schlesinger}},\ and\ \bibinfo {author} {\bibfnamefont {E.~M.}\ \bibnamefont
  {Mueller}},\ }\href@noop {} {\emph {\bibinfo {title} {Alloy phase
  diagrams}}}\ (\bibinfo  {publisher} {Asm International},\ \bibinfo {year}
  {2016})\BibitemShut {NoStop}%
\bibitem [{\citenamefont {Sch{\"o}n}\ and\ \citenamefont
  {Jansen}(2009)}]{SchonJansen2009}%
  \BibitemOpen
  \bibfield  {author} {\bibinfo {author} {\bibfnamefont {J.}~\bibnamefont
  {Sch{\"o}n}}\ and\ \bibinfo {author} {\bibfnamefont {M.}~\bibnamefont
  {Jansen}},\ }\href@noop {} {\bibfield  {journal} {\bibinfo  {journal}
  {International Journal of Materials Research}\ }\textbf {\bibinfo {volume}
  {100}},\ \bibinfo {pages} {135} (\bibinfo {year} {2009})}\BibitemShut
  {NoStop}%
\bibitem [{\citenamefont {Enoki}\ \emph {et~al.}(2023)\citenamefont {Enoki},
  \citenamefont {Minamoto}, \citenamefont {Ohnuma}, \citenamefont {Abe},\ and\
  \citenamefont {Ohtani}}]{ISIJReview2023}%
  \BibitemOpen
  \bibfield  {author} {\bibinfo {author} {\bibfnamefont {M.}~\bibnamefont
  {Enoki}}, \bibinfo {author} {\bibfnamefont {S.}~\bibnamefont {Minamoto}},
  \bibinfo {author} {\bibfnamefont {I.}~\bibnamefont {Ohnuma}}, \bibinfo
  {author} {\bibfnamefont {T.}~\bibnamefont {Abe}},\ and\ \bibinfo {author}
  {\bibfnamefont {H.}~\bibnamefont {Ohtani}},\ }\href@noop {} {\bibfield
  {journal} {\bibinfo  {journal} {ISIJ International}\ }\textbf {\bibinfo
  {volume} {63}},\ \bibinfo {pages} {407} (\bibinfo {year} {2023})}\BibitemShut
  {NoStop}%
\bibitem [{\citenamefont {Campbell}(2012)}]{campbell2012phase}%
  \BibitemOpen
  \bibinfo {editor} {\bibfnamefont {F.~C.}\ \bibnamefont {Campbell}},\ ed.,\
  \href@noop {} {\emph {\bibinfo {title} {Phase Diagrams: Understanding the
  Basics}}}\ (\bibinfo  {publisher} {ASM International},\ \bibinfo {address}
  {Materials Park, OH},\ \bibinfo {year} {2012})\ \bibinfo {note} {introductory
  reference on alloy phase diagrams, thermodynamics, and phase
  fields}\BibitemShut {NoStop}%
\bibitem [{\citenamefont {Zhao}(2007)}]{zhao2007methods}%
  \BibitemOpen
  \bibfield  {author} {\bibinfo {author} {\bibfnamefont {J.-C.}\ \bibnamefont
  {Zhao}},\ }\href@noop {} {\emph {\bibinfo {title} {Methods for Phase Diagram
  Determination}}}\ (\bibinfo  {publisher} {Elsevier},\ \bibinfo {year}
  {2007})\BibitemShut {NoStop}%
\bibitem [{\citenamefont {Lukas}\ \emph {et~al.}(2007)\citenamefont {Lukas},
  \citenamefont {Fries},\ and\ \citenamefont
  {Sundman}}]{Lukas2007ComputationalThermodynamics}%
  \BibitemOpen
  \bibfield  {author} {\bibinfo {author} {\bibfnamefont {H.}~\bibnamefont
  {Lukas}}, \bibinfo {author} {\bibfnamefont {S.~G.}\ \bibnamefont {Fries}},\
  and\ \bibinfo {author} {\bibfnamefont {B.}~\bibnamefont {Sundman}},\
  }\href@noop {} {\emph {\bibinfo {title} {Computational thermodynamics: the
  Calphad method}}}\ (\bibinfo  {publisher} {Cambridge university press},\
  \bibinfo {year} {2007})\BibitemShut {NoStop}%
\bibitem [{\citenamefont {Saunders}\ and\ \citenamefont
  {Miodownik}(1998)}]{saunders1998calphad}%
  \BibitemOpen
  \bibfield  {author} {\bibinfo {author} {\bibfnamefont {N.}~\bibnamefont
  {Saunders}}\ and\ \bibinfo {author} {\bibfnamefont {A.~P.}\ \bibnamefont
  {Miodownik}},\ }\href@noop {} {\emph {\bibinfo {title} {CALPHAD (calculation
  of phase diagrams): a comprehensive guide}}},\ Vol.~\bibinfo {volume} {1}\
  (\bibinfo  {publisher} {Elsevier},\ \bibinfo {year} {1998})\BibitemShut
  {NoStop}%
\bibitem [{\citenamefont {Zipoli}\ \emph {et~al.}(2022)\citenamefont {Zipoli},
  \citenamefont {Asahara},\ and\ \citenamefont {Kanehira}}]{iecr2022_ml_pd}%
  \BibitemOpen
  \bibfield  {author} {\bibinfo {author} {\bibfnamefont {F.}~\bibnamefont
  {Zipoli}}, \bibinfo {author} {\bibfnamefont {A.}~\bibnamefont {Asahara}},\
  and\ \bibinfo {author} {\bibfnamefont {T.}~\bibnamefont {Kanehira}},\
  }\href@noop {} {\bibfield  {journal} {\bibinfo  {journal} {Ind. Eng. Chem.
  Res.}\ }\textbf {\bibinfo {volume} {61}},\ \bibinfo {pages} {7412} (\bibinfo
  {year} {2022})}\BibitemShut {NoStop}%
\bibitem [{\citenamefont {Wang}\ \emph {et~al.}(2024)\citenamefont {Wang},
  \citenamefont {Wang}, \citenamefont {Zhang}, \citenamefont {Li},
  \citenamefont {Zhou},\ and\ \citenamefont {Sun}}]{wang2024machine}%
  \BibitemOpen
  \bibfield  {author} {\bibinfo {author} {\bibfnamefont {G.}~\bibnamefont
  {Wang}}, \bibinfo {author} {\bibfnamefont {C.}~\bibnamefont {Wang}}, \bibinfo
  {author} {\bibfnamefont {X.}~\bibnamefont {Zhang}}, \bibinfo {author}
  {\bibfnamefont {Z.}~\bibnamefont {Li}}, \bibinfo {author} {\bibfnamefont
  {J.}~\bibnamefont {Zhou}},\ and\ \bibinfo {author} {\bibfnamefont
  {Z.}~\bibnamefont {Sun}},\ }\href@noop {} {\bibfield  {journal} {\bibinfo
  {journal} {Iscience}\ }\textbf {\bibinfo {volume} {27}} (\bibinfo {year}
  {2024})}\BibitemShut {NoStop}%
\bibitem [{\citenamefont {Zhu}\ \emph {et~al.}(2025)\citenamefont {Zhu},
  \citenamefont {Sariturk},\ and\ \citenamefont
  {Arroyave}}]{mlip_phaseforge_2025}%
  \BibitemOpen
  \bibfield  {author} {\bibinfo {author} {\bibfnamefont {S.}~\bibnamefont
  {Zhu}}, \bibinfo {author} {\bibfnamefont {D.}~\bibnamefont {Sariturk}},\ and\
  \bibinfo {author} {\bibfnamefont {R.}~\bibnamefont {Arroyave}},\ }\href@noop
  {} {\bibfield  {journal} {\bibinfo  {journal} {npj Comput. Mater.}\ }\textbf
  {\bibinfo {volume} {11}},\ \bibinfo {pages} {340} (\bibinfo {year}
  {2025})}\BibitemShut {NoStop}%
\bibitem [{\citenamefont {van~de Walle}\ and\ \citenamefont
  {Ceder}(2002)}]{van2002automating}%
  \BibitemOpen
  \bibfield  {author} {\bibinfo {author} {\bibfnamefont {A.}~\bibnamefont
  {van~de Walle}}\ and\ \bibinfo {author} {\bibfnamefont {G.}~\bibnamefont
  {Ceder}},\ }\href@noop {} {\bibfield  {journal} {\bibinfo  {journal} {J.
  Phase Equilibria}\ }\textbf {\bibinfo {volume} {23}},\ \bibinfo {pages} {348}
  (\bibinfo {year} {2002})}\BibitemShut {NoStop}%
\bibitem [{\citenamefont {Liu}\ \emph {et~al.}(2021)\citenamefont {Liu},
  \citenamefont {Esteban-Manzanares},\ and\ \citenamefont
  {LLorca}}]{liu2021first}%
  \BibitemOpen
  \bibfield  {author} {\bibinfo {author} {\bibfnamefont {S.}~\bibnamefont
  {Liu}}, \bibinfo {author} {\bibfnamefont {G.}~\bibnamefont
  {Esteban-Manzanares}},\ and\ \bibinfo {author} {\bibfnamefont
  {J.}~\bibnamefont {LLorca}},\ }\href@noop {} {\bibfield  {journal} {\bibinfo
  {journal} {Metallurgical and Materials Transactions A}\ }\textbf {\bibinfo
  {volume} {52}},\ \bibinfo {pages} {4675} (\bibinfo {year}
  {2021})}\BibitemShut {NoStop}%
\bibitem [{\citenamefont {Alfe}\ \emph {et~al.}(2002)\citenamefont {Alfe},
  \citenamefont {Gillan},\ and\ \citenamefont {Price}}]{alfe2002ab}%
  \BibitemOpen
  \bibfield  {author} {\bibinfo {author} {\bibfnamefont {D.}~\bibnamefont
  {Alfe}}, \bibinfo {author} {\bibfnamefont {M.}~\bibnamefont {Gillan}},\ and\
  \bibinfo {author} {\bibfnamefont {G.}~\bibnamefont {Price}},\ }\href@noop {}
  {\bibfield  {journal} {\bibinfo  {journal} {The Journal of chemical physics}\
  }\textbf {\bibinfo {volume} {116}},\ \bibinfo {pages} {7127} (\bibinfo {year}
  {2002})}\BibitemShut {NoStop}%
\bibitem [{\citenamefont {Alf{\`e}}\ \emph
  {et~al.}(2002{\natexlab{a}})\citenamefont {Alf{\`e}}, \citenamefont
  {Gillan},\ and\ \citenamefont {Price}}]{alfe2002composition}%
  \BibitemOpen
  \bibfield  {author} {\bibinfo {author} {\bibfnamefont {D.}~\bibnamefont
  {Alf{\`e}}}, \bibinfo {author} {\bibfnamefont {M.}~\bibnamefont {Gillan}},\
  and\ \bibinfo {author} {\bibfnamefont {G.~D.}\ \bibnamefont {Price}},\
  }\href@noop {} {\bibfield  {journal} {\bibinfo  {journal} {Earth and
  Planetary Science Letters}\ }\textbf {\bibinfo {volume} {195}},\ \bibinfo
  {pages} {91} (\bibinfo {year} {2002}{\natexlab{a}})}\BibitemShut {NoStop}%
\bibitem [{\citenamefont {Chipman}(1972)}]{chipman1972thermodynamics}%
  \BibitemOpen
  \bibfield  {author} {\bibinfo {author} {\bibfnamefont {J.}~\bibnamefont
  {Chipman}},\ }\href@noop {} {\bibfield  {journal} {\bibinfo  {journal}
  {Metallurgical Transactions}\ }\textbf {\bibinfo {volume} {3}},\ \bibinfo
  {pages} {55} (\bibinfo {year} {1972})}\BibitemShut {NoStop}%
\bibitem [{\citenamefont {Hirose}\ \emph {et~al.}(2021)\citenamefont {Hirose},
  \citenamefont {Wood},\ and\ \citenamefont {Vo{\v{c}}adlo}}]{hirose2021light}%
  \BibitemOpen
  \bibfield  {author} {\bibinfo {author} {\bibfnamefont {K.}~\bibnamefont
  {Hirose}}, \bibinfo {author} {\bibfnamefont {B.}~\bibnamefont {Wood}},\ and\
  \bibinfo {author} {\bibfnamefont {L.}~\bibnamefont {Vo{\v{c}}adlo}},\
  }\href@noop {} {\bibfield  {journal} {\bibinfo  {journal} {Nature Reviews
  Earth \& Environment}\ }\textbf {\bibinfo {volume} {2}},\ \bibinfo {pages}
  {645} (\bibinfo {year} {2021})}\BibitemShut {NoStop}%
\bibitem [{\citenamefont {Zhang}\ \emph {et~al.}(2022)\citenamefont {Zhang},
  \citenamefont {Cs{\'a}nyi}, \citenamefont {Alf{\`e}}, \citenamefont {Zhang},
  \citenamefont {Li},\ and\ \citenamefont {Liu}}]{zhang2022free}%
  \BibitemOpen
  \bibfield  {author} {\bibinfo {author} {\bibfnamefont {Z.}~\bibnamefont
  {Zhang}}, \bibinfo {author} {\bibfnamefont {G.}~\bibnamefont {Cs{\'a}nyi}},
  \bibinfo {author} {\bibfnamefont {D.}~\bibnamefont {Alf{\`e}}}, \bibinfo
  {author} {\bibfnamefont {Y.}~\bibnamefont {Zhang}}, \bibinfo {author}
  {\bibfnamefont {J.}~\bibnamefont {Li}},\ and\ \bibinfo {author}
  {\bibfnamefont {J.}~\bibnamefont {Liu}},\ }\href@noop {} {\bibfield
  {journal} {\bibinfo  {journal} {Geophysical Research Letters}\ }\textbf
  {\bibinfo {volume} {49}},\ \bibinfo {pages} {e2021GL096749} (\bibinfo {year}
  {2022})}\BibitemShut {NoStop}%
\bibitem [{\citenamefont {Sharma}\ \emph {et~al.}(2025)\citenamefont {Sharma},
  \citenamefont {Mehta},\ and\ \citenamefont {Alf{\`e}}}]{sharma2025ab}%
  \BibitemOpen
  \bibfield  {author} {\bibinfo {author} {\bibfnamefont {S.~B.}\ \bibnamefont
  {Sharma}}, \bibinfo {author} {\bibfnamefont {S.}~\bibnamefont {Mehta}},\ and\
  \bibinfo {author} {\bibfnamefont {D.}~\bibnamefont {Alf{\`e}}},\ }\href@noop
  {} {\bibfield  {journal} {\bibinfo  {journal} {The Journal of Chemical
  Physics}\ }\textbf {\bibinfo {volume} {162}} (\bibinfo {year}
  {2025})}\BibitemShut {NoStop}%
\bibitem [{\citenamefont {Mendelev}\ \emph {et~al.}(2009)\citenamefont
  {Mendelev}, \citenamefont {Asta}, \citenamefont {Rahman},\ and\ \citenamefont
  {Hoyt}}]{AlMg-theory}%
  \BibitemOpen
  \bibfield  {author} {\bibinfo {author} {\bibfnamefont {M.}~\bibnamefont
  {Mendelev}}, \bibinfo {author} {\bibfnamefont {M.}~\bibnamefont {Asta}},
  \bibinfo {author} {\bibfnamefont {M.}~\bibnamefont {Rahman}},\ and\ \bibinfo
  {author} {\bibfnamefont {J.}~\bibnamefont {Hoyt}},\ }\href@noop {} {\bibfield
   {journal} {\bibinfo  {journal} {Phil. Mag.}\ }\textbf {\bibinfo {volume}
  {89}},\ \bibinfo {pages} {3269} (\bibinfo {year} {2009})}\BibitemShut
  {NoStop}%
\bibitem [{\citenamefont {Liu}\ \emph {et~al.}(2013)\citenamefont {Liu},
  \citenamefont {Roven}, \citenamefont {Murashkin}, \citenamefont {Valiev},
  \citenamefont {Kilmametov}, \citenamefont {Zhang},\ and\ \citenamefont
  {Yu}}]{liu2013structure}%
  \BibitemOpen
  \bibfield  {author} {\bibinfo {author} {\bibfnamefont {M.~P.}\ \bibnamefont
  {Liu}}, \bibinfo {author} {\bibfnamefont {H.~J.}\ \bibnamefont {Roven}},
  \bibinfo {author} {\bibfnamefont {M.~Y.}\ \bibnamefont {Murashkin}}, \bibinfo
  {author} {\bibfnamefont {R.~Z.}\ \bibnamefont {Valiev}}, \bibinfo {author}
  {\bibfnamefont {A.}~\bibnamefont {Kilmametov}}, \bibinfo {author}
  {\bibfnamefont {Z.}~\bibnamefont {Zhang}},\ and\ \bibinfo {author}
  {\bibfnamefont {Y.}~\bibnamefont {Yu}},\ }\href@noop {} {\bibfield  {journal}
  {\bibinfo  {journal} {J. Mater. Sci.}\ }\textbf {\bibinfo {volume} {48}},\
  \bibinfo {pages} {4681} (\bibinfo {year} {2013})}\BibitemShut {NoStop}%
\bibitem [{\citenamefont {Murray}(1982)}]{murray1982mg}%
  \BibitemOpen
  \bibfield  {author} {\bibinfo {author} {\bibfnamefont {J.~L.}\ \bibnamefont
  {Murray}},\ }\href@noop {} {\bibfield  {journal} {\bibinfo  {journal} {J.
  Phase Equilib.}\ }\textbf {\bibinfo {volume} {3}},\ \bibinfo {pages} {60}
  (\bibinfo {year} {1982})}\BibitemShut {NoStop}%
\bibitem [{\citenamefont {Daw}\ and\ \citenamefont
  {Baskes}(1983)}]{daw1983semiempirical}%
  \BibitemOpen
  \bibfield  {author} {\bibinfo {author} {\bibfnamefont {M.~S.}\ \bibnamefont
  {Daw}}\ and\ \bibinfo {author} {\bibfnamefont {M.~I.}\ \bibnamefont
  {Baskes}},\ }\href@noop {} {\bibfield  {journal} {\bibinfo  {journal} {Phys.
  Rev. Lett.}\ }\textbf {\bibinfo {volume} {50}},\ \bibinfo {pages} {1285}
  (\bibinfo {year} {1983})}\BibitemShut {NoStop}%
\bibitem [{\citenamefont {Allen}\ and\ \citenamefont
  {Tildesley}(2017)}]{AllenTildesley2017}%
  \BibitemOpen
  \bibfield  {author} {\bibinfo {author} {\bibfnamefont {M.~P.}\ \bibnamefont
  {Allen}}\ and\ \bibinfo {author} {\bibfnamefont {D.~J.}\ \bibnamefont
  {Tildesley}},\ }\href@noop {} {\emph {\bibinfo {title} {Computer simulation
  of liquids}}}\ (\bibinfo  {publisher} {Oxford university press},\ \bibinfo
  {year} {2017})\BibitemShut {NoStop}%
\bibitem [{\citenamefont {Alf{\`e}}\ \emph
  {et~al.}(2002{\natexlab{b}})\citenamefont {Alf{\`e}}, \citenamefont
  {Gillan},\ and\ \citenamefont {Price}}]{alfe2002complementary}%
  \BibitemOpen
  \bibfield  {author} {\bibinfo {author} {\bibfnamefont {D.}~\bibnamefont
  {Alf{\`e}}}, \bibinfo {author} {\bibfnamefont {M.}~\bibnamefont {Gillan}},\
  and\ \bibinfo {author} {\bibfnamefont {G.}~\bibnamefont {Price}},\
  }\href@noop {} {\bibfield  {journal} {\bibinfo  {journal} {J. Chem. Phys.}\
  }\textbf {\bibinfo {volume} {116}},\ \bibinfo {pages} {6170} (\bibinfo {year}
  {2002}{\natexlab{b}})}\BibitemShut {NoStop}%
\bibitem [{\citenamefont {Mehta}\ \emph {et~al.}(2006)\citenamefont {Mehta},
  \citenamefont {Price},\ and\ \citenamefont {Alf{\`e}}}]{mehta2006ab}%
  \BibitemOpen
  \bibfield  {author} {\bibinfo {author} {\bibfnamefont {S.}~\bibnamefont
  {Mehta}}, \bibinfo {author} {\bibfnamefont {G.}~\bibnamefont {Price}},\ and\
  \bibinfo {author} {\bibfnamefont {D.}~\bibnamefont {Alf{\`e}}},\ }\href@noop
  {} {\bibfield  {journal} {\bibinfo  {journal} {The Journal of chemical
  physics}\ }\textbf {\bibinfo {volume} {125}} (\bibinfo {year}
  {2006})}\BibitemShut {NoStop}%
\bibitem [{\citenamefont {Alf{\`e}}(2009)}]{alfe2009phon}%
  \BibitemOpen
  \bibfield  {author} {\bibinfo {author} {\bibfnamefont {D.}~\bibnamefont
  {Alf{\`e}}},\ }\href@noop {} {\bibfield  {journal} {\bibinfo  {journal}
  {Computer Physics Communications}\ }\textbf {\bibinfo {volume} {180}},\
  \bibinfo {pages} {2622} (\bibinfo {year} {2009})}\BibitemShut {NoStop}%
\bibitem [{\citenamefont {Kresse}\ and\ \citenamefont
  {Furthm{"u}ller}(1996)}]{kresse1996efficient}%
  \BibitemOpen
  \bibfield  {author} {\bibinfo {author} {\bibfnamefont {G.}~\bibnamefont
  {Kresse}}\ and\ \bibinfo {author} {\bibfnamefont {J.}~\bibnamefont
  {Furthm{"u}ller}},\ }\href@noop {} {\bibfield  {journal} {\bibinfo  {journal}
  {Physical Review B}\ }\textbf {\bibinfo {volume} {54}},\ \bibinfo {pages}
  {11169} (\bibinfo {year} {1996})}\BibitemShut {NoStop}%
\bibitem [{\citenamefont {Kresse}\ and\ \citenamefont
  {Joubert}(1999)}]{kresse1999ultrasoft}%
  \BibitemOpen
  \bibfield  {author} {\bibinfo {author} {\bibfnamefont {G.}~\bibnamefont
  {Kresse}}\ and\ \bibinfo {author} {\bibfnamefont {D.}~\bibnamefont
  {Joubert}},\ }\href@noop {} {\bibfield  {journal} {\bibinfo  {journal}
  {Physical review b}\ }\textbf {\bibinfo {volume} {59}},\ \bibinfo {pages}
  {1758} (\bibinfo {year} {1999})}\BibitemShut {NoStop}%
\bibitem [{\citenamefont {Bl{"o}chl}(1994)}]{blochl1994projector}%
  \BibitemOpen
  \bibfield  {author} {\bibinfo {author} {\bibfnamefont {P.~E.}\ \bibnamefont
  {Bl{"o}chl}},\ }\href@noop {} {\bibfield  {journal} {\bibinfo  {journal}
  {Physical Review B}\ }\textbf {\bibinfo {volume} {50}},\ \bibinfo {pages}
  {17953} (\bibinfo {year} {1994})}\BibitemShut {NoStop}%
\bibitem [{\citenamefont {Perdew}\ \emph {et~al.}(1996)\citenamefont {Perdew},
  \citenamefont {Burke},\ and\ \citenamefont
  {Ernzerhof}}]{perdew1996generalized}%
  \BibitemOpen
  \bibfield  {author} {\bibinfo {author} {\bibfnamefont {J.~P.}\ \bibnamefont
  {Perdew}}, \bibinfo {author} {\bibfnamefont {K.}~\bibnamefont {Burke}},\ and\
  \bibinfo {author} {\bibfnamefont {M.}~\bibnamefont {Ernzerhof}},\ }\href@noop
  {} {\bibfield  {journal} {\bibinfo  {journal} {Physical review letters}\
  }\textbf {\bibinfo {volume} {77}},\ \bibinfo {pages} {3865} (\bibinfo {year}
  {1996})}\BibitemShut {NoStop}%
\bibitem [{\citenamefont {Mermin}(1965)}]{mermin1965thermal}%
  \BibitemOpen
  \bibfield  {author} {\bibinfo {author} {\bibfnamefont {N.~D.}\ \bibnamefont
  {Mermin}},\ }\href@noop {} {\bibfield  {journal} {\bibinfo  {journal}
  {Physical Review}\ }\textbf {\bibinfo {volume} {137}},\ \bibinfo {pages}
  {A1441} (\bibinfo {year} {1965})}\BibitemShut {NoStop}%
\bibitem [{\citenamefont {Nos{\'e}}(1984)}]{nose1984molecular}%
  \BibitemOpen
  \bibfield  {author} {\bibinfo {author} {\bibfnamefont {S.}~\bibnamefont
  {Nos{\'e}}},\ }\href@noop {} {\bibfield  {journal} {\bibinfo  {journal}
  {Molecular physics}\ }\textbf {\bibinfo {volume} {52}},\ \bibinfo {pages}
  {255} (\bibinfo {year} {1984})}\BibitemShut {NoStop}%
\bibitem [{\citenamefont {Monkhorst}\ and\ \citenamefont
  {Pack}(1976)}]{monkhorst1976special}%
  \BibitemOpen
  \bibfield  {author} {\bibinfo {author} {\bibfnamefont {H.~J.}\ \bibnamefont
  {Monkhorst}}\ and\ \bibinfo {author} {\bibfnamefont {J.~D.}\ \bibnamefont
  {Pack}},\ }\href@noop {} {\bibfield  {journal} {\bibinfo  {journal} {Physical
  review B}\ }\textbf {\bibinfo {volume} {13}},\ \bibinfo {pages} {5188}
  (\bibinfo {year} {1976})}\BibitemShut {NoStop}%
\bibitem [{\citenamefont {Vo{\v{c}}adlo}\ and\ \citenamefont
  {Alf{\`e}}(2002)}]{vocadlo2002ab}%
  \BibitemOpen
  \bibfield  {author} {\bibinfo {author} {\bibfnamefont {L.}~\bibnamefont
  {Vo{\v{c}}adlo}}\ and\ \bibinfo {author} {\bibfnamefont {D.}~\bibnamefont
  {Alf{\`e}}},\ }\href@noop {} {\bibfield  {journal} {\bibinfo  {journal}
  {Physical Review B}\ }\textbf {\bibinfo {volume} {65}},\ \bibinfo {pages}
  {214105} (\bibinfo {year} {2002})}\BibitemShut {NoStop}%
\bibitem [{\citenamefont {Polsin}\ \emph {et~al.}(2018)\citenamefont {Polsin},
  \citenamefont {Fratanduono}, \citenamefont {Rygg}, \citenamefont {Lazicki},
  \citenamefont {Smith}, \citenamefont {Eggert}, \citenamefont {Gregor},
  \citenamefont {Henderson}, \citenamefont {Gong}, \citenamefont {Delettrez}
  \emph {et~al.}}]{polsin2018x}%
  \BibitemOpen
  \bibfield  {author} {\bibinfo {author} {\bibfnamefont {D.}~\bibnamefont
  {Polsin}}, \bibinfo {author} {\bibfnamefont {D.}~\bibnamefont {Fratanduono}},
  \bibinfo {author} {\bibfnamefont {J.}~\bibnamefont {Rygg}}, \bibinfo {author}
  {\bibfnamefont {A.}~\bibnamefont {Lazicki}}, \bibinfo {author} {\bibfnamefont
  {R.}~\bibnamefont {Smith}}, \bibinfo {author} {\bibfnamefont
  {J.}~\bibnamefont {Eggert}}, \bibinfo {author} {\bibfnamefont
  {M.}~\bibnamefont {Gregor}}, \bibinfo {author} {\bibfnamefont
  {B.}~\bibnamefont {Henderson}}, \bibinfo {author} {\bibfnamefont
  {X.}~\bibnamefont {Gong}}, \bibinfo {author} {\bibfnamefont {J.}~\bibnamefont
  {Delettrez}}, \emph {et~al.},\ }\href@noop {} {\bibfield  {journal} {\bibinfo
   {journal} {Physics of Plasmas}\ }\textbf {\bibinfo {volume} {25}} (\bibinfo
  {year} {2018})}\BibitemShut {NoStop}%
\bibitem [{\citenamefont {Akahama}\ \emph {et~al.}(2006)\citenamefont
  {Akahama}, \citenamefont {Nishimura}, \citenamefont {Kinoshita},
  \citenamefont {Kawamura},\ and\ \citenamefont
  {Ohishi}}]{akahama2006evidence}%
  \BibitemOpen
  \bibfield  {author} {\bibinfo {author} {\bibfnamefont {Y.}~\bibnamefont
  {Akahama}}, \bibinfo {author} {\bibfnamefont {M.}~\bibnamefont {Nishimura}},
  \bibinfo {author} {\bibfnamefont {K.}~\bibnamefont {Kinoshita}}, \bibinfo
  {author} {\bibfnamefont {H.}~\bibnamefont {Kawamura}},\ and\ \bibinfo
  {author} {\bibfnamefont {Y.}~\bibnamefont {Ohishi}},\ }\href@noop {}
  {\bibfield  {journal} {\bibinfo  {journal} {Physical review letters}\
  }\textbf {\bibinfo {volume} {96}},\ \bibinfo {pages} {045505} (\bibinfo
  {year} {2006})}\BibitemShut {NoStop}%
\bibitem [{\citenamefont {Kudasov}\ \emph {et~al.}(2013)\citenamefont
  {Kudasov}, \citenamefont {Surdin}, \citenamefont {Korshunov}, \citenamefont
  {Pavlov}, \citenamefont {Frolova},\ and\ \citenamefont
  {Kuzin}}]{kudasov2013lattice}%
  \BibitemOpen
  \bibfield  {author} {\bibinfo {author} {\bibfnamefont {Y.~B.}\ \bibnamefont
  {Kudasov}}, \bibinfo {author} {\bibfnamefont {O.}~\bibnamefont {Surdin}},
  \bibinfo {author} {\bibfnamefont {A.}~\bibnamefont {Korshunov}}, \bibinfo
  {author} {\bibfnamefont {V.}~\bibnamefont {Pavlov}}, \bibinfo {author}
  {\bibfnamefont {N.}~\bibnamefont {Frolova}},\ and\ \bibinfo {author}
  {\bibfnamefont {R.}~\bibnamefont {Kuzin}},\ }\href@noop {} {\bibfield
  {journal} {\bibinfo  {journal} {Journal of Experimental and Theoretical
  Physics}\ }\textbf {\bibinfo {volume} {117}},\ \bibinfo {pages} {664}
  (\bibinfo {year} {2013})}\BibitemShut {NoStop}%
\bibitem [{\citenamefont {Bouchet}\ \emph {et~al.}(2009)\citenamefont
  {Bouchet}, \citenamefont {Bottin}, \citenamefont {Jomard},\ and\
  \citenamefont {Z{\'e}rah}}]{bouchet2009melting}%
  \BibitemOpen
  \bibfield  {author} {\bibinfo {author} {\bibfnamefont {J.}~\bibnamefont
  {Bouchet}}, \bibinfo {author} {\bibfnamefont {F.}~\bibnamefont {Bottin}},
  \bibinfo {author} {\bibfnamefont {G.}~\bibnamefont {Jomard}},\ and\ \bibinfo
  {author} {\bibfnamefont {G.}~\bibnamefont {Z{\'e}rah}},\ }\href@noop {}
  {\bibfield  {journal} {\bibinfo  {journal} {Physical Review B—Condensed
  Matter and Materials Physics}\ }\textbf {\bibinfo {volume} {80}},\ \bibinfo
  {pages} {094102} (\bibinfo {year} {2009})}\BibitemShut {NoStop}%
\bibitem [{\citenamefont {Boehler}\ and\ \citenamefont
  {Ross}(1997)}]{boehler1997melting}%
  \BibitemOpen
  \bibfield  {author} {\bibinfo {author} {\bibfnamefont {R.}~\bibnamefont
  {Boehler}}\ and\ \bibinfo {author} {\bibfnamefont {M.}~\bibnamefont {Ross}},\
  }\href@noop {} {\bibfield  {journal} {\bibinfo  {journal} {Earth Planet. Sci.
  Lett.}\ }\textbf {\bibinfo {volume} {153}},\ \bibinfo {pages} {223} (\bibinfo
  {year} {1997})}\BibitemShut {NoStop}%
\bibitem [{\citenamefont {H{\"a}nstr{\"o}m}\ and\ \citenamefont
  {Lazor}(2000)}]{hanstrom2000high}%
  \BibitemOpen
  \bibfield  {author} {\bibinfo {author} {\bibfnamefont {A.}~\bibnamefont
  {H{\"a}nstr{\"o}m}}\ and\ \bibinfo {author} {\bibfnamefont {P.}~\bibnamefont
  {Lazor}},\ }\href@noop {} {\bibfield  {journal} {\bibinfo  {journal} {J.
  Alloys Compd.}\ }\textbf {\bibinfo {volume} {305}},\ \bibinfo {pages} {209}
  (\bibinfo {year} {2000})}\BibitemShut {NoStop}%
\bibitem [{\citenamefont {Errandonea}(2010)}]{errandonea2010melting}%
  \BibitemOpen
  \bibfield  {author} {\bibinfo {author} {\bibfnamefont {D.}~\bibnamefont
  {Errandonea}},\ }\href@noop {} {\bibfield  {journal} {\bibinfo  {journal} {J.
  Appl. Phys.}\ }\textbf {\bibinfo {volume} {108}} (\bibinfo {year}
  {2010})}\BibitemShut {NoStop}%
\bibitem [{\citenamefont {Homan}\ \emph {et~al.}(1984)\citenamefont {Homan},
  \citenamefont {MacCrone},\ and\ \citenamefont {Whalley}}]{homan1984high}%
  \BibitemOpen
  \bibfield  {author} {\bibinfo {author} {\bibfnamefont {C.}~\bibnamefont
  {Homan}}, \bibinfo {author} {\bibfnamefont {R.}~\bibnamefont {MacCrone}},\
  and\ \bibinfo {author} {\bibfnamefont {E.}~\bibnamefont {Whalley}},\
  }\href@noop {} {\bibfield  {journal} {\bibinfo  {journal} {Parts I, II, III}\
  } (\bibinfo {year} {1984})}\BibitemShut {NoStop}%
\bibitem [{\citenamefont {Olijnyk}\ and\ \citenamefont
  {Holzapfel}(1985)}]{olijnyk1985high}%
  \BibitemOpen
  \bibfield  {author} {\bibinfo {author} {\bibfnamefont {H.}~\bibnamefont
  {Olijnyk}}\ and\ \bibinfo {author} {\bibfnamefont {W.}~\bibnamefont
  {Holzapfel}},\ }\href@noop {} {\bibfield  {journal} {\bibinfo  {journal}
  {Physical Review B}\ }\textbf {\bibinfo {volume} {31}},\ \bibinfo {pages}
  {4682} (\bibinfo {year} {1985})}\BibitemShut {NoStop}%
\bibitem [{\citenamefont {Stinton}\ \emph {et~al.}(2014)\citenamefont
  {Stinton}, \citenamefont {MacLeod}, \citenamefont {Cynn}, \citenamefont
  {Errandonea}, \citenamefont {Evans}, \citenamefont {Proctor}, \citenamefont
  {Meng},\ and\ \citenamefont {McMahon}}]{stinton2014equation}%
  \BibitemOpen
  \bibfield  {author} {\bibinfo {author} {\bibfnamefont {G.~W.}\ \bibnamefont
  {Stinton}}, \bibinfo {author} {\bibfnamefont {S.~G.}\ \bibnamefont
  {MacLeod}}, \bibinfo {author} {\bibfnamefont {H.}~\bibnamefont {Cynn}},
  \bibinfo {author} {\bibfnamefont {D.}~\bibnamefont {Errandonea}}, \bibinfo
  {author} {\bibfnamefont {W.~J.}\ \bibnamefont {Evans}}, \bibinfo {author}
  {\bibfnamefont {J.~E.}\ \bibnamefont {Proctor}}, \bibinfo {author}
  {\bibfnamefont {Y.}~\bibnamefont {Meng}},\ and\ \bibinfo {author}
  {\bibfnamefont {M.~I.}\ \bibnamefont {McMahon}},\ }\href@noop {} {\bibfield
  {journal} {\bibinfo  {journal} {Physical Review B}\ }\textbf {\bibinfo
  {volume} {90}},\ \bibinfo {pages} {134105} (\bibinfo {year}
  {2014})}\BibitemShut {NoStop}%
\bibitem [{\citenamefont {Cui}\ \emph {et~al.}(2022)\citenamefont {Cui},
  \citenamefont {Xian}, \citenamefont {Liu}, \citenamefont {Tian},
  \citenamefont {Gao},\ and\ \citenamefont {Song}}]{cui2022melting}%
  \BibitemOpen
  \bibfield  {author} {\bibinfo {author} {\bibfnamefont {C.}~\bibnamefont
  {Cui}}, \bibinfo {author} {\bibfnamefont {J.}~\bibnamefont {Xian}}, \bibinfo
  {author} {\bibfnamefont {H.}~\bibnamefont {Liu}}, \bibinfo {author}
  {\bibfnamefont {F.}~\bibnamefont {Tian}}, \bibinfo {author} {\bibfnamefont
  {X.}~\bibnamefont {Gao}},\ and\ \bibinfo {author} {\bibfnamefont
  {H.}~\bibnamefont {Song}},\ }\href@noop {} {\bibfield  {journal} {\bibinfo
  {journal} {Journal of Applied Physics}\ }\textbf {\bibinfo {volume} {131}}
  (\bibinfo {year} {2022})}\BibitemShut {NoStop}%
\bibitem [{\citenamefont {Fletcher}\ \emph {et~al.}(2025)\citenamefont
  {Fletcher}, \citenamefont {Bart{\'o}k},\ and\ \citenamefont
  {P{\'a}rtay}}]{fletcher2025autonomous}%
  \BibitemOpen
  \bibfield  {author} {\bibinfo {author} {\bibfnamefont {V.~G.}\ \bibnamefont
  {Fletcher}}, \bibinfo {author} {\bibfnamefont {A.~P.}\ \bibnamefont
  {Bart{\'o}k}},\ and\ \bibinfo {author} {\bibfnamefont {L.~B.}\ \bibnamefont
  {P{\'a}rtay}},\ }\href@noop {} {\bibfield  {journal} {\bibinfo  {journal}
  {npj Computational Materials}\ } (\bibinfo {year} {2025})}\BibitemShut
  {NoStop}%
\bibitem [{\citenamefont {Hong}\ and\ \citenamefont {Van
  De~Walle}(2019)}]{hong2019reentrant}%
  \BibitemOpen
  \bibfield  {author} {\bibinfo {author} {\bibfnamefont {Q.-J.}\ \bibnamefont
  {Hong}}\ and\ \bibinfo {author} {\bibfnamefont {A.}~\bibnamefont {Van
  De~Walle}},\ }\href@noop {} {\bibfield  {journal} {\bibinfo  {journal}
  {Physical Review B}\ }\textbf {\bibinfo {volume} {100}},\ \bibinfo {pages}
  {140102} (\bibinfo {year} {2019})}\BibitemShut {NoStop}%
\bibitem [{\citenamefont {Sansonetti}\ and\ \citenamefont
  {Martin}(2005)}]{sansonetti2005handbook}%
  \BibitemOpen
  \bibfield  {author} {\bibinfo {author} {\bibfnamefont {J.~E.}\ \bibnamefont
  {Sansonetti}}\ and\ \bibinfo {author} {\bibfnamefont {W.~C.}\ \bibnamefont
  {Martin}},\ }\href@noop {} {\bibfield  {journal} {\bibinfo  {journal}
  {Journal of physical and chemical reference data}\ }\textbf {\bibinfo
  {volume} {34}},\ \bibinfo {pages} {1559} (\bibinfo {year}
  {2005})}\BibitemShut {NoStop}%
\bibitem [{\citenamefont {Cannon}(1974)}]{cannon1974behavior}%
  \BibitemOpen
  \bibfield  {author} {\bibinfo {author} {\bibfnamefont {J.~F.}\ \bibnamefont
  {Cannon}},\ }\href@noop {} {\bibfield  {journal} {\bibinfo  {journal}
  {Journal of Physical and Chemical Reference Data}\ }\textbf {\bibinfo
  {volume} {3}},\ \bibinfo {pages} {781} (\bibinfo {year} {1974})}\BibitemShut
  {NoStop}%
\bibitem [{\citenamefont {Chase~Jr}\ \emph {et~al.}(1985)\citenamefont
  {Chase~Jr}, \citenamefont {Davies}, \citenamefont {Downey~Jr}, \citenamefont
  {Frurip}, \citenamefont {McDonald},\ and\ \citenamefont
  {Syverud}}]{chase1985janaf}%
  \BibitemOpen
  \bibfield  {author} {\bibinfo {author} {\bibfnamefont {M.}~\bibnamefont
  {Chase~Jr}}, \bibinfo {author} {\bibfnamefont {C.}~\bibnamefont {Davies}},
  \bibinfo {author} {\bibfnamefont {J.}~\bibnamefont {Downey~Jr}}, \bibinfo
  {author} {\bibfnamefont {D.}~\bibnamefont {Frurip}}, \bibinfo {author}
  {\bibfnamefont {R.}~\bibnamefont {McDonald}},\ and\ \bibinfo {author}
  {\bibfnamefont {A.}~\bibnamefont {Syverud}},\ }\href@noop {} {\bibfield
  {journal} {\bibinfo  {journal} {Journal of physical and chemical reference
  data}\ }\textbf {\bibinfo {volume} {14}},\ \bibinfo {pages} {927} (\bibinfo
  {year} {1985})}\BibitemShut {NoStop}%
\bibitem [{\citenamefont {Courac}\ \emph {et~al.}(2020)\citenamefont {Courac},
  \citenamefont {Le~Godec}, \citenamefont {Solozhenko}, \citenamefont
  {Guignot},\ and\ \citenamefont {Crichton}}]{courac2020thermoelastic}%
  \BibitemOpen
  \bibfield  {author} {\bibinfo {author} {\bibfnamefont {A.}~\bibnamefont
  {Courac}}, \bibinfo {author} {\bibfnamefont {Y.}~\bibnamefont {Le~Godec}},
  \bibinfo {author} {\bibfnamefont {V.~L.}\ \bibnamefont {Solozhenko}},
  \bibinfo {author} {\bibfnamefont {N.}~\bibnamefont {Guignot}},\ and\ \bibinfo
  {author} {\bibfnamefont {W.~A.}\ \bibnamefont {Crichton}},\ }\href@noop {}
  {\bibfield  {journal} {\bibinfo  {journal} {Journal of Applied Physics}\
  }\textbf {\bibinfo {volume} {127}} (\bibinfo {year} {2020})}\BibitemShut
  {NoStop}%
\bibitem [{\citenamefont {Chase}\ \emph {et~al.}(1998)\citenamefont {Chase}
  \emph {et~al.}}]{chase1998nist}%
  \BibitemOpen
  \bibfield  {author} {\bibinfo {author} {\bibfnamefont {M.~W.}\ \bibnamefont
  {Chase}} \emph {et~al.},\ }\href@noop {} {\bibfield  {journal} {\bibinfo
  {journal} {Journal of physical and chemical reference data}\ }\textbf
  {\bibinfo {volume} {28}},\ \bibinfo {pages} {1951} (\bibinfo {year}
  {1998})}\BibitemShut {NoStop}%
\end{thebibliography}%

\onecolumngrid
\begin{center}
UK Ministry of Defence \textcopyright\ Crown owned copyright 2026/AWE
\end{center}
\twocolumngrid

\end{document}